\documentclass[fleqn,usenatbib]{mnras}

\usepackage{newtxtext,newtxmath}

\usepackage[T1]{fontenc}
\usepackage{ae,aecompl}
\usepackage{footnote}
\usepackage{soul}
\usepackage{CJK}

\usepackage{graphicx}    
\usepackage{amsmath}    
\usepackage{amssymb}    
\usepackage{array, longtable}

\newcommand{\angstrom}{\text{\normalfont\AA}}
\newcommand{\Msun}{M$_{\odot}$}
\newcommand{\teff}{$T_\mathrm{eff}$}

\newcommand{\lppr}{\stackrel{<}{\scriptstyle \sim}}
\newcommand{\lappr}{\raisebox{-0.4ex}{$\lppr$}}

\title[IR-excess WDs in the {\it Gaia} 100 pc sample]{Infrared-excess
  white dwarfs in the {\it Gaia} 100 pc sample}

  \author[Rebassa-Mansergas et
  al.]{A. Rebassa-Mansergas$^{1,2}$\thanks{E-mail:
    alberto.rebassa@upc.edu}, E. Solano$^{3,4}$, S. Xu$^{5}$, C. Rodrigo$^{3,4}$,\newauthor
  F. M. Jim\'enez-Esteban$^{3,4}$,  S. Torres$^{1,2}$ \\
$^{1}$ Departament de F\'{\i}sica, Universitat Polit\`{e}cnica de Catalunya, c/Esteve Terrades 5, 08860 Castelldefels, Spain\\
$^{2}$ Institut d'Estudis Espacials de Catalunya, Ed. Nexus-201, c/Gran Capit\`a 2-4, 08034 Barcelona, Spain\\
$^{3}$Departmento de Astrof\'{\i}sica, Centro de Astrobiolog\'{\i}a (CSIC-INTA), ESAC Campus, Camino Bajo del Castillo s/n,\\
E-28692 Villanueva de la Ca\~nada, Madrid, Spain\\
$^{4}$ Spanish Virtual Observatory, E-28692 Villanueva de la Ca\~nada, Madrid, Spain\\
$^{5}$ Gemini Observatory, 670 N. A'ohoku Place, Hilo, HI 96720
}

\date{Accepted XXX. Received YYY; in original form ZZZ}

\pubyear{2019}
\begin{document}
\label{firstpage}
\pagerange{\pageref{firstpage}--\pageref{lastpage}}
\maketitle

\begin{abstract}
We analyse the 100\,pc {\it Gaia} white dwarf volume-limited sample by
means  of VOSA  (Virtual Observatory  SED  Analyser) with  the aim  of
identifying candidates  for displaying infrared excesses.   Our search
focuses  on the  study of  the spectral  energy distribution  (SED) of
3,733  white  dwarfs with  reliable  infrared  photometry and  $G_{\rm
  BP}-G_{\rm RP}$  colours below 0.8 mag,  a sample which seems  to be
nearly  representative of  the  overall white  dwarf population.   Our
search  results  in  77  selected  candidates, 52  of  which  are  new
identifications.  For each target we apply a two-component SED fitting
implemented in VOSA  to derive the effective temperatures  of both the
white  dwarf  and the  object  causing  the  excess.  We  calculate  a
fraction  of infrared-excess  white dwarfs  due to  the presence  of a
circumstellar  disk  of  1.6$\pm$0.2\%,  a value  which  increases  to
2.6$\pm$0.3\%  if we  take  into account  incompleteness issues.   Our
results are in  agreement with the drop in the  percentage of infrared
excess detections for cool  ($<$8,000\,K) and hot ($>$20,000\,K) white
dwarfs obtained  in previous analyses.   The fraction of  white dwarfs
with brown dwarf companions we derive is $\simeq$0.1--0.2\%.
\end{abstract}




\section{Introduction}

Low-  and  intermediate-mass  main sequence  stars  ($M\,\lappr\,8\sim
11\,M_{\sun}$)   end  their   lives   as  white   dwarfs  (WDs;   e.g.
\citealt{Siess2007}).  WDs are hence  the most common stellar remnants
and are one of  the most common objects in the  Galaxy. Given that WDs
are  compact  objects  and  nuclear reactions  have  ceased  in  their
interiors,  their  structure  is  supported by  the  pressure  of  the
degenerate electrons  in their  cores. The energy  reservoir available
from previous evolutionary phases  is contained within this degenerate
core  and radiated  away  through a  thin  envelope of  non-degenerate
matter  following a  moderately well-understood  cooling process  (see
e.g. the review by \citealt{Althaus2010a}  and reference therein for a
thorough discussion of this issue).  This envelope is generally formed
by  an upper  layer of  hydrogen of  10$^{-2}$-10$^{-4}$\,\Msun and  a
lower  layer  of   helium  of  10$^{-15}$-10$^{-5}$\,\Msun  \citep[see
  e.g.][]{Castanheira2008,  Tremblay2008}.  Due  to  the high  surface
gravity acting  on WD atmospheres,  the heavier elements  sink towards
the deep interiors.  Hence, the optical spectra of the majority of WDs
show Balmer absorption lines  typical of hydrogen-rich atmospheres, or
helium   absorption   lines   if   this   hydrogen   layer   is   lost
\citep{Bergeron2011,  Koester2015}.  However,  25-50 per  cent of  WDs
show    heavy    elements    apart   from    hydrogen    and    helium
\citep{Zuckerman2003, Koester2014}. These WDs  are referred to as DAZs
and DBZs, respectively, or DZs if only metal lines are observed. It is
of  vital  importance  to  understand how  these  metals  reached  the
atmosphere of those WDs.

Planets and minor planets located a few  AUs away from a host star are
expected to survive the giant phases  once the star evolves out of the
main sequence  and becomes  a WD \citep{Burleigh2002,  Jura2008}. This
implies  the orbits  of  these planets  expand,  a rearrangement  that
causes instability to the system.  This perturbation may cause some of
the surviving minor  planets to enter into the tidal  radius of the WD
and, as a consequence, to  be disrupted and accreted \citep{Debes2002,
  Debes2012}   during   a  process   that   can   last  a   few   Gyrs
\citep{Bonsor2011,  Veras2013}. The  accretion  of planetary  material
leads to the enrichment of heavy elements in the atmosphere of the WD,
explaining the identification  of metal transitions in  the spectra of
such DZ  WDs.  An  additional observational feature  that arises  as a
consequence of the disruption of a minor planet is a dust and/or a gas
disk within the  tidal radius of the  WD \citep{Gaensicke2006, Xu2012,
  Dennihy2018}. It has been observed that  some of these dusty WDs are
dynamically active \citep{xu+jura2014} and  recently, transits from an
actively disintegrating  asteroid have  been discovered for  the first
time  around a  dusty  WD  \citep{Vandergurg2015}.  The  observational
feature of  dust disks around  WDs is  the detection of  infrared (IR)
excess  \citep[e.g.][]{Zuckerman1987}.  Apart  from  a few  exceptions
\citep{Xu2015, Wilson2019},  the great  majority of dusty  WDs display
also traces  of heavy  elements in their  atmospheres.  The  dust disk
occurrence  is about  1--4\% for  WDs \citep{Barber2012,  Rocchetto15,
  Wilson2019}. For  less than  0.5\% of the  cases, the  excess arises
from   the  existence   of   a  sub-stellar   brown  dwarf   companion
\citep{Farihi2004,  Farihi2005}.    It  is  also  possible   that  the
IR-excess arises  from the presence  of low-mass star  companions that
are  outshined in  the optical  by  the flux  of a  relatively hot  WD
\citep{Rebassa2010, Badenes2013, Rebassa2016}.

The  presence of  metals in  the atmosphere  of a  WD provides  unique
information   about  the   composition   of   the  accreted   material
\citep{Klein2010,  Gaensicke2012,  Xu2014,  Hollands2018b},  which  is
found to be chemically  Earth-like \citep{wilson2016}.  Pollution rich
in water  \citep{Farihi2013, Raddi2015} as well  as water/ice-rich and
volatile-rich  (C  \&  N)   \citep{Xu2017}  has  also  been  observed.
Unfortunately, the  current number of \emph{confirmed}  WDs by Spitzer
observations displaying infrared excess due to a circumstellar disk is
just 35 (see  the recent review by  \citealt{Farihi2016} and reference
therein), which makes  it difficult to characterise  the properties of
extreme  planetary  systems.   In  this work  we  aim  at  identifying
additional  IR-excess  WD  candidates,  in  particular  those  with  a
circumstellar disk, by analysing  the most complete and volume-limited
sample of WDs  to date, identified thanks to the  data provided by the
second release of the {\it Gaia} mission.

\begin{figure*}
    \includegraphics[width=\columnwidth]{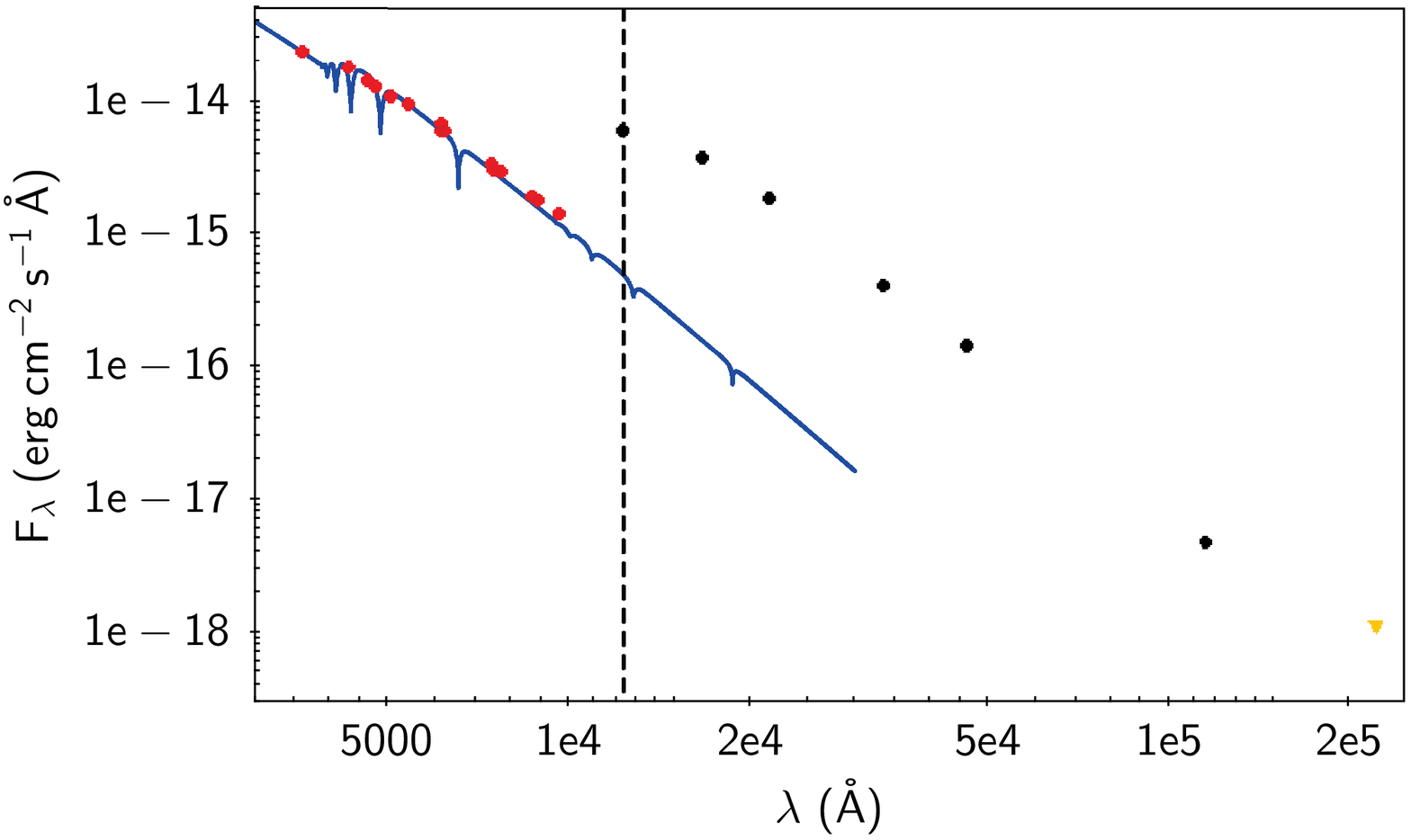}
    \hspace{2cm}
    \includegraphics[width=0.6\columnwidth]{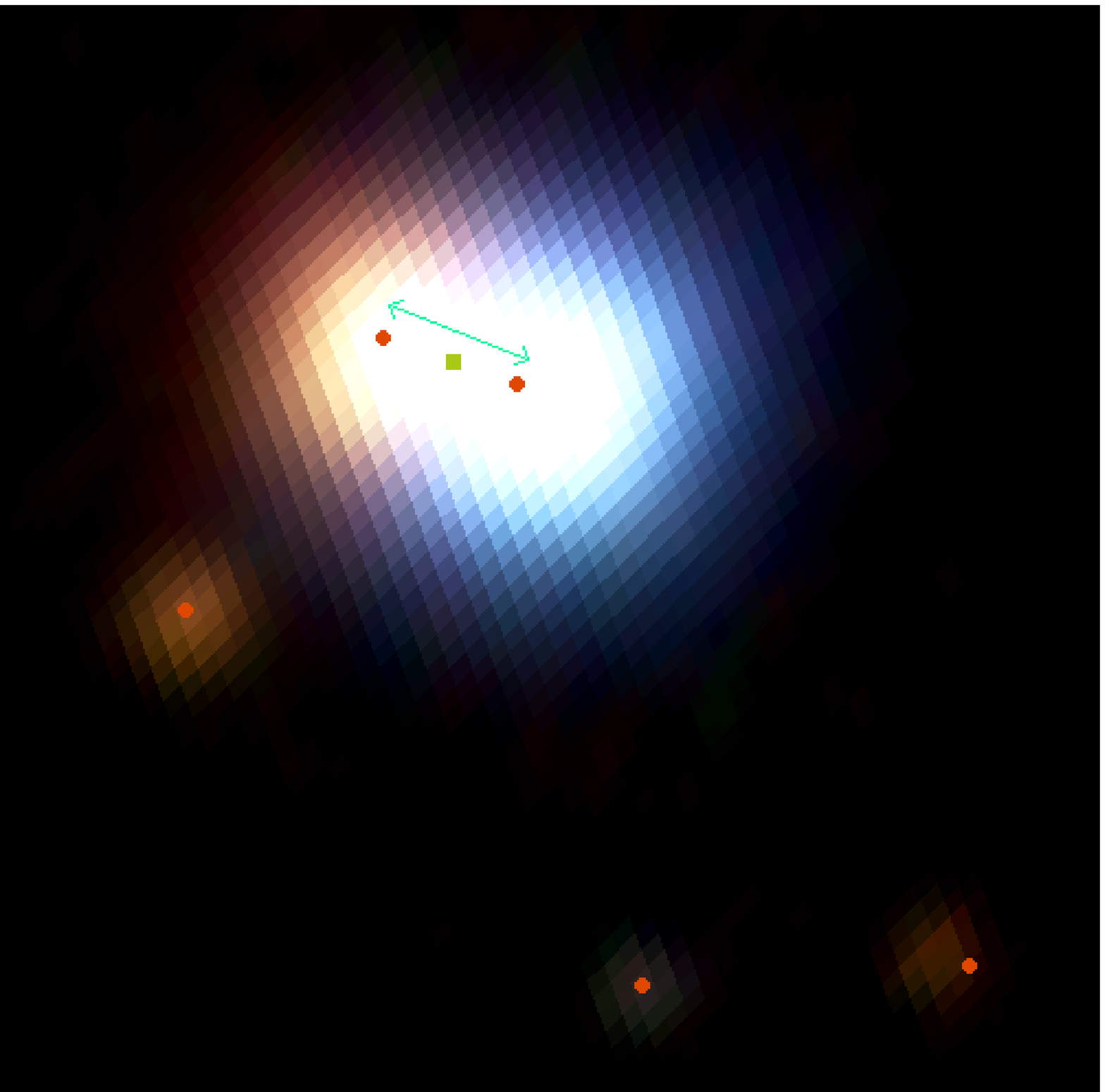}
     \includegraphics[width=\columnwidth]{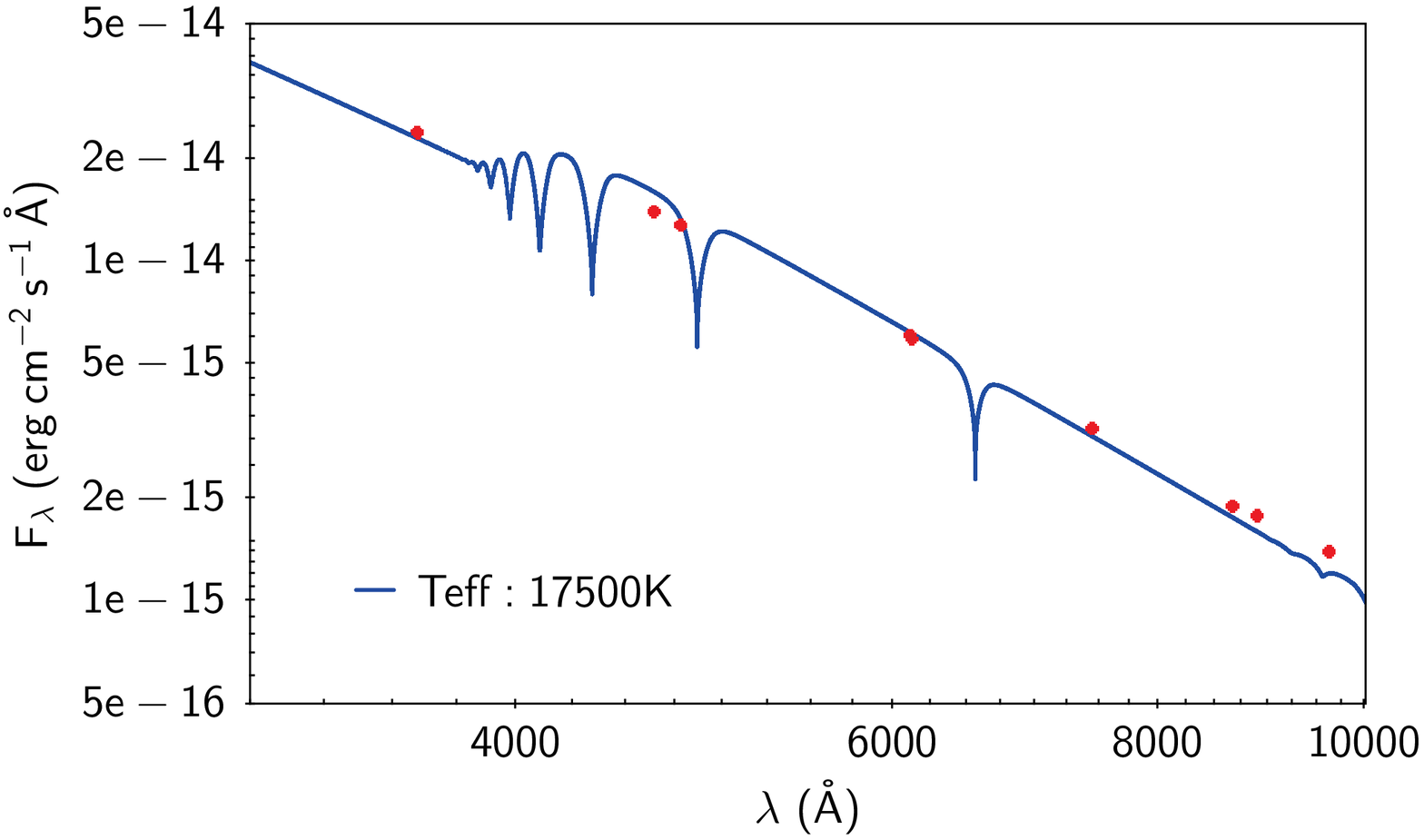}
      \includegraphics[width=\columnwidth]{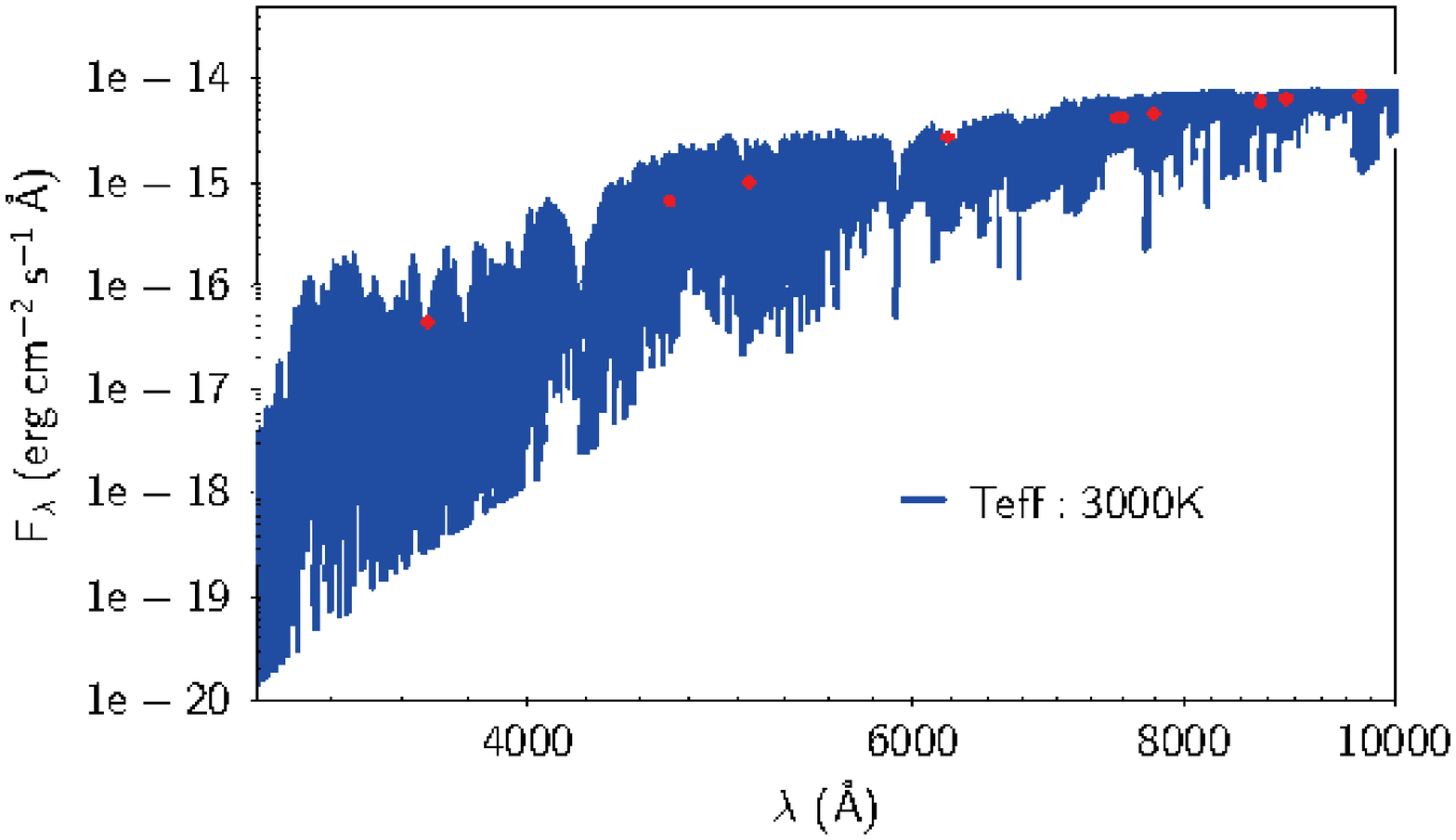}
    \caption{Top  left:   Example  of  a  composite   SED  of  GaiaID:
      1075673567146680576 and GaiaID:1075673567146680704  (IDs 11A and
      11B  in  Table\,\ref{t:co-moving}).   The   abrupt  jump  at  IR
      wavelengths indicates that more  than one object is contributing
      to  the  stellar  flux.   The vertical  dashed  line  marks  the
      wavelength at which  VOSA detects {\it excess}  (understood as a
      significant change  in the slope  and/or a clear  deviation from
      the  photospheric  flux predicted  by  the  model).  The  yellow
      inverted  triangle indicates  that the  photometric value  is an
      upper  limit.   Top  right:  SDSS image  showing  two  partially
      resolved objects.  Red bullets  represent SDSS sources while the
      green square indicates the position of the single WISE catalogue
      entry  for the  two sources.   Bottom  panels: The  SEDs of  the
      individual objects built using photometry from surveys of higher
      spatial  resolution (SDSS,  PanStarrs) along  with the  best fit
      models and their associated effective temperatures.  Photometric
      errors are too small to be seen.}
    \label{fig:bin}
\end{figure*}

\section{Identification of infrared-excess White Dwarf candidates}
\label{Iden}

The data provided by the {\it  Gaia} satellite through its Second Data
Release has  allowed identifying unprecedented samples  of both single
and  binary WDs  \citep[e.g.][]{Badry2018,  Gentile2019},  as well  as
compiling the  largest and  most complete volume-limited  catalogue of
such objects to date within 100 pc \citep{Jimenez-Esteban18}. In order
to   identify   WDs  with   IR   excess   we   took  the   sample   of
\citet{Jimenez-Esteban18} of  8,555 objects  at less  than 100  pc and
with $G_{\rm BP}-G_{\rm  RP}$ colours below 0.8 mag  available at "The
SVO       archive      of       White      Dwarfs       from      {\it
  Gaia}"\footnote{http://svo2.cab.inta-csic.es/vocats/v2/wdw/}.   This
colour cut is equivalent to excluding  WDs cooler than 6,000\,K and it
is  required due  to the  expected large  contamination of  non-WDs at
lower  effective  temperatures  (see  \citealt{Jimenez-Esteban18}  for
details).  We analysed the Spectral Energy Distributions (SEDs) of the
8,555           WDs            taking           advantage           of
VOSA\footnote{http://svo.cab.inta-csic.es/theory/vosa/}       (Virtual
Observatory  SED  Analyser;  \citealt{Bayo08}).   VOSA  is  a  Virtual
Observatory tool that allows building the SEDs of thousands of objects
in  an automated  way from  a large  number of  photometric catalogues
ranging  from the  ultraviolet  to the  infrared.   VOSA compares  the
photometric data with different  collections of theoretical models and
determines  which model  best reproduces  the observed  data following
different   statistical   approaches.    Physical   parameters   (e.g.
effective temperatures,  luminosities) are then estimated  for each WD
from  the  model that  best  fits  the  data.   VOSA also  allows  the
identification of  IR-excess in  the SED  and estimates  the effective
temperature and  luminosity of  the source  causing the  excess, which
together with  the Stefan-Boltzmann  equation yields also  its radius.
Note that  the calculated  radius has physical  meaning only  when the
source of excess is a companion.

\begin{table*}
\caption{List  of  the  26  co-moving   WD+M  systems  found  in  this
  work. Coordinates, parallaxes and proper motions are from {\it Gaia}
  DR2. The  object names  are obtained from  Simbad and  the effective
  temperatures    have     been    estimated    as     described    in
  Sect\,\ref{Iden}. $\ast$ in the second last column indicates that we
  were not  able to estimate  \teff\, due to  the lack of  enough good
  photometric  points to  run the  SED  fitting. In  these cases,  the
  spectral  types for  the cool  components were  estimated using  the
  $G$-$G_\mathrm{RP}$   colour  and   the   calibration  provided   by
  \citet{Pecaut2013}.}
\setlength{\tabcolsep}{0.8ex}
\begin{tabular}{cccccccccc} 
\hline
ID & Gaia ID & RA(ICRS) & DEC(ICRS) & Name & Parallax & PMRA & PMDE & \teff & comment\\
   & DR2     & deg  & deg  &      & mas      & mas/yr & mas/yr & K\\
\hline
1A &  396370097820352256 & 22.4283 & 42.4715 & GD 13      &    11.88$\pm$0.09 &   94.49$\pm$0.12 & -45.59$\pm$0.11 & 22,000 &(1)\\
1B & 396370093526060672 & 22.4271 & 42.4704 &   & 11.39$\pm$0.17 &   92.30$\pm$0.21 &-45.30$\pm$0.24 & 2,900&\\
2A&5119295082016649472 & 37.3367 & -24.4348 & & 13.05$\pm$0.05 & 91.76$\pm$0.1 & -94.85$\pm$0.10 & 15,500 &(1)\\
2B& 5119295082017370368 & 37.3359 & -24.4350& & 13.14$\pm$0.06 &  89.27$\pm$0.11 & -98.94$\pm$0.12 &3,200&\\
3A & 18493721155296768 & 39.6289 & 6.6333 & PG 0235+064 & 16.89$\pm$0.06 & -12.77$\pm$0.09 & -90.99$\pm$0.08 & 13,000 &(1)\\
3B & 18493721155296640 & 39.6283 & 6.6353 & 2MASS J02383078+0638071 & 16.98$\pm$0.07 & -11.22$\pm$0.11 & -87.69$\pm$0.10 & 3200&\\
4A & 16426811093797760 & 51.5877 & 11.5158 & & 16.38$\pm$0.06 & -49.14$\pm$0.12 & -36.06$\pm$0.08 & 17,500 &(1)\\
4B & 16426806798453632 & 51.5884 & 11.5149 &[ZEH2003] RX J0326.3+1131 3 & 16.12$\pm$0.08 & -44.65$\pm$0.18 & -35.08$\pm$0.12 & M2V$\ast$ &\\
5A & 166587938734739456 & 64.1762  & 32.1891 & SDSS J041642.29+321120.5 & 10.13$\pm$0.09 & 0.86$\pm$0.17 & -83.38$\pm$0.13& 8250 & (2) \\
5B & 166587938734739328 & 64.1769 &  32.1891 &  & 10.42$\pm$0.16 & 2.16$\pm$0,32 & -80.71$\pm$0.23 & 3100 & \\
6A & 4811421896276768128 & 73.3033 & -44.3944 & & 11.13$\pm$0.04 & 26.87$\pm$0.07 & 23.22$\pm$0.08 & WD$\ast$ &(1)\\
6B & 4811421896275732480 & 73.3026 & -44.3937 & & 11.17$\pm$0.08 & 25.32$\pm$0.16 & 23.01$\pm$0.18 & M4-5V$\ast$ &\\
7A & 201854258801563520 & 74.1309 &  41.5220 &    &  10.13$\pm$0.09 &   -36.68$\pm$0.16 &  -106.43$\pm$0.11 & 9,500 &(1)\\
7B & 201854258801564288 & 74.1293 & 41.5218 & &10.09$\pm$0.07 &   -35.79$\pm$0.13 & -106.98$\pm$0.09 &  3,300 &\\
8A & 3439162768415866112 & 96.5553 & 32.2198& & 12.58$\pm$0.06 & -17.35$\pm$0.11 & -38.63$\pm$0.1& 11,000 &(1)\\
8B & 3439162768415865600 & 96.5566 & 32.2202& & 12.71$\pm$0.11& -19.45$\pm$0.20& -37.31$\pm$0.18 & 3,200 &\\
9A& 5598661329740179712 & 116.5578 & -30.4313& & 14.49$\pm$0.05 & 59.43$\pm$0.07 & -117.24$\pm$0.08 & 17,500 &(1)\\
9B & 5598661329740179584 & 116.5575 & -30.4305 & & 14.39$\pm$0.06 & 59.54$\pm$0.09 & -117.89$\pm$0.1 & 3,100 &\\
10A & 3842126835031738368 & 137.9023 & -0.2159 & & 16.80$\pm$0.19 & 59.78$\pm$0.29 & -36.23$\pm$0.25 & 6,250 &(1)\\
10B & 3842126835031738496 & 137.9020 & -0.2144 & & 16.66$\pm$0.21 & 56.93$\pm$0.36 & -35.10$\pm$0.29 & 2,800 &\\
11A & 1075673567146680576 & 170.0952 & 72.8795 &   & 11.50$\pm$0.06 &  -63.50$\pm$0.10 &   5.56$\pm$0.09 & 15,750&(1)\\ 
11B& 1075673567146680704 & 170.0922 & 72.8795 &  &  11.60$\pm$0.06 &  -64.37$\pm$0.11&   3.08$\pm$0.09 & 3,600 &\\
12A & 1692021543289085184 & 190.5089 & 75.1460 & PG\,1240+754 & 12.42$\pm$ 0.05 & -201.85$\pm$0.08 & -35.09$\pm$0.07 & WD$\ast$ & (3) \\
12B & 1692021543289084672 & 190.5142 & 75.1450 &  G\,255-B18B & 11.98$\pm$0.06 & -199.77$\pm$0.11 & -32.49$\pm$0.09 & 3200 & \\
13A & 1552488776081383040 & 204.0067 &   48.4793 &  GD 325 & 27.09$\pm$0.04 & -134.10$\pm$0.04 & -42.98$\pm$0.05 & 17,500&(1)\\ 
13B & 1552488776081383168 & 204.0079 &  48.4796 &   & 27.06$\pm$0.06 &  -127.55$\pm$0.07 &  -47.55$\pm$0.09  & 3,000&\\
14A &  1604422214954487168 & 216.6841 &  50.1066 & CBS 268 &   15.28$\pm$0.04 & -10.60$\pm$0.06 & -82.72$\pm$0.06 & 15,250& (4)\\
14B & 1604422283673964160 & 216.6830 & 50.1093 &  & 15.20$\pm$0.04 &  -11.70$\pm$0.06 & -84.83$\pm$0.06 &  3,300&\\
15A & 1276054682231244160 & 225.4845 & 30.3831 & PG 1459+306 & 15.00$\pm$0.04 & -43.15$\pm$0.04 & 52.77$\pm$0.06 &19,250&(1)\\   
15B & 1276054677930790272 & 225.4851 & 30.3842 &             &  14.94$\pm$0.03& -37.31$\pm$0.03 & 54.16$\pm$0.04 & 3,500&\\
16A & 1643551566043342848 & 240.7042 & 67.4912& & 12.28$\pm$0.08 & -25.19$\pm$0.15 & 6.98$\pm$0.13 & 8,000 &(1)\\
16B & 1643551566043342592 & 240.7044 & 67.4898 & & 12.32$\pm$0.03 & -24.36$\pm$0.06 & 4.07$\pm$0.05 & 3,500 &\\
17A & 4457170451083163392 & 242.7226 &  11.7313 & PG 1608+119 & 11.59$\pm$0.06 &  36.49$\pm$0.07 &  -12.99$\pm$0.06  & 20,000&(1)\\
17B & 4457170446785639424  & 242.7218 &   11.7316 & &  11.58$\pm$0.05 & 33.30$\pm$0.06 & -13.97$\pm$0.05 & 3,500&\\
18A &  1300356053864952064 & 251.6571 &  25.3068 &  &12.32$\pm$0.05 &  -44.96$\pm$0.07 & 2.40$\pm$0.11 & 12,750&(1)\\
18B &  1300356809779196288 & 251.6561 &  25.3070 &  &12.25$\pm$0.06 &  -43.78$\pm$0.09 &  3.70$\pm$0.13 & 3,100 &\\
19A& 4360643809885839232 & 257.3352& -7.8785& & 13.82$\pm$0.08&-37.94$\pm$0.13& -112.47$\pm$0.09 & 17,750 &(1)\\
19B & 4360643809885838976 & 257.3362 & -7.8790& & 13.91$\pm$0.06 & -41.21$\pm$0.10& -110.21$\pm$0.07 & 3,400&\\
20A & 4366961260100103680& 260.7060& -2.8049& & 12.34$\pm$0.09& -17.64$\pm$0.16 & -29.71$\pm$0.13 & 7,250 &(4)\\
20B & 4366961260100103552 & 260.7054 & -2.8054& & 11.87$\pm$0.07 & -19.20$\pm$0.11 & -29.08$\pm$0.09 & 3,200& \\
21A & 5803547624984209792 & 266.1464 & -72.9932 & & 11.43$\pm$0.05 & 14.44$\pm$0.06 & 17.53$\pm$0.08 & WD$\ast$ & (4) \\
21B & 5803547624984209664 & 266.1488 & -72.9932 & & 11.36$\pm$0.10 & 16.45$\pm$0.11 & 14.49$\pm$0.15 & M5V$\ast$ &  \\
22A & 2103614787618232192 & 284.3691 &  40.5932 & &  15.00$\pm$0.03 & -23.46$\pm$0.05 &  41.47$\pm$0.05  & 17,500&(1)\\
22B & 2103614787618232448  & 284.3697 &  40.5938 & KIC 5342558& 15.12$\pm$0.05 &  -26.09$\pm$0.09 & 42.79$\pm$0.09 & 3,100&\\
23A & 1768730586908531712 & 330.3371 &  15.0917 & & 10.68$\pm$0.06 &  -30.32$\pm$0.11 & -89.21$\pm$0.10 & 14,250&(1)\\
23B & 1768730586908531840  & 330.3375 & 15.0928 & & 10.63$\pm$0.04 &  -30.56$\pm$0.07 &  -88.05$\pm$0.06 & 3,500&\\
24A & 2811484217573797248 & 347.3354 & 11.5747& & 14.33$\pm$0.20 & 23.92$\pm$0.35 & -62.61$\pm$0.25 & 7,000 &(1)\\
24B &2811484217572663168 &  347.3361 & 11.5753 &  & 14.37$\pm$0.17 & 24.35$\pm$0.31 & -65.31$\pm$0.21 & 2,900 &\\
25A & 6393502099375879552 & 348.2538 & -64.3321 & & 12.71$\pm$0.05 & 179.05$\pm$0.07 & 23.96$\pm$0.08 & WD$\ast$ &(4)\\
25B & 6393502099376815744 & 348.2521 & -64.3322 & & 13.36$\pm$0.24 & 184.32$\pm$0.34 & 19.37$\pm$0.36 & M3-4V$\ast$ &\\
26A& 1999127510441929600 & 356.2808 & 58.2209 & & 10.22$\pm$0.08 & 28.87$\pm$0.11 & 5.83$\pm$0.08 & 11,750 &(4)\\
26B & 1999127510436274048 & 356.2794 & 58.2205 & & 10.54$\pm$0.39 & 29.84$\pm$0.61 & 5.69$\pm$0.42& 2,800 &\\
\hline
\end{tabular}
\label{t:co-moving}
\begin{list}{}{} 
\item[]
(1) \citet{Badry2018};
(2) \citet{Ren2014};
(3) Simbad;
(4) This work.
\end{list}
\end{table*}

In  this paper  we made  use of  the following  photometric catalogues
available   at  VOSA:   GALEX   \citep{Bianchi00},   {\it  Gaia}   DR2
\citep{Brown18},    SDSS   DR12    \citep{Alam15},   Pan-STARRS    DR1
\citep{Chambers16}, the Dark Energy Survey (DES) \citep{DESC16}, 2MASS
PSC     \citep{Skrutskie06},     VISTA     \citep{Cross12},     UKIDSS
\citep{Hewett06}, and WISE \citep{Wright10}. Additionally, we made use
of  the   Spitzer  \   Enhanced  Imaging  Products   (SEIP)  catalogue
\citep{Wu10},  which is  presently  not included  in  VOSA.  To  avoid
potential  mismatches  we  used  the  {\it  Gaia}  proper  motions  to
calculate the corresponding {\it Gaia} coordinates at the J2000 epoch,
used by all other surveys considered\footnote{Ideally, for high proper
  motion  objects  one  would  require to  work  out  the  \emph{Gaia}
  coordinates  at  the  exact  epoch  of  observations  of  the  other
  different surveys.  However, VOSA uses  a search radius of 5", which
  means a  WD needs a proper  motion higher than 330  mas/year to move
  more than  5" in 15  years (J2000 to J2015).   Less than 1\%  of the
  8,555 WDs within 100pc from \citet{Jimenez-Esteban18} have such high
  proper   motions.   Therefore,   not  calculating   the  \emph{Gaia}
  coordinates at the exact epoch  of observations of the other surveys
  has a very low impact.}.

From the original  list, we filtered out objects with  less than three
reliable IR ($>$12,000\AA) photometric points in their SEDs.  Reliable
photometry  implies data  not  affected by  contamination from  nearby
sources, artifacts or quality flag issues  (Qflg $\neq$ U in the J and
H  bands  for 2MASS;  ccf=0  and  qph=A/B in  the  W2  band for  WISE;
ppErrBits<256  for both  VISTA and  UKIDSS).  This  resulted in  3,733
selected {\it Gaia} WDs.  In order to explore the possibility that our
selected  sample is  representative of  the overall  WD population  we
compared  the   corresponding  WD   effective  temperature   and  mass
distributions  (see Section\,\ref{s-disc}  for  details  on how  these
parameters  are  derived)  to  those arising  from  the  catalogue  of
\emph{Gaia} DA WDs within 20\,pc of \citet{Hollands2018}.  This is not
only a volume-limited and complete sample, but also all 20 pc WDs have
available    effective    temperature   and    mass    determinations.
Kolmogorov-Smirnov  (KS)   tests  yield  probabilities   of  10$^{-5}$
(4.4$\sigma$; effective temperature) and  0.55 (0.6$\sigma$; mass) for
our  and the  20\,pc \emph{Gaia}  samples to  be drawn  from the  same
parent population. If we exclude  cool ($<$6,000K) WDs from the 20\,pc
sample,  the effective  temperature KS  probability increases  to 0.16
(1.4$\sigma$). We  thus conclude there  are no strong  indications for
our  sample not  to be  representative of  the overall  WD population,
except at  effective temperature values  under 6,000\,K.  The  lack of
such cool WDs in our sample is not surprising since these are excluded
by our imposed $G_{\rm BP}-G_{\rm RP}<0.8$ colour cut.

The observational SEDs of the 3,733 selected objects were compared to
the  hydrogen-rich  WD  collection  of  theoretical  model  atmosphere
spectra      of     \citet{Koester10}      (see     Sect.\,4.1      in
\citealt{Jimenez-Esteban18}  for a  detailed description  of the  main
characteristics of  this grid of  models) to identify IR  excesses. To
that  end, VOSA  first executes  an  iterative algorithm  which is  an
extension  of  the method  described  in  \citet{Lada06}. Starting  at
$\lambda \geq 21500 \angstrom$, VOSA  computes the slope of the linear
regression  of   the  observational   SED  in  a   log\,$\nu  F_{\nu}$
vs.  log\,$\nu$  diagram.  This  slope is  recomputed  by  adding  new
infrared photometric points at every step.  If, in any of these steps,
the  slope  becomes  significantly  smaller  ($<2.56$)  than  the  one
expected from a  stellar photospheric emission, VOSA  flags the object
as potentially affected by IR  excess and photometric points at longer
wavelengths are not taken into account  in the SED fitting process for
deriving the WD effective temperature and luminosity.

Once the SED fitting is  completed, VOSA performs a further refinement
of the IR excess estimation  by comparing, for each photometric point,
the observational flux  to the synthetic flux obtained  from the model
that best  fits the  data. Significant ($>$\,3$\sigma$)  deviations in
the observational flux are flagged by VOSA as potential IR excesses. A
detailed description  of how VOSA  manages the infrared excess  can be
found                   in                  the                   VOSA
documentation\footnote{https://bit.ly/2KRCv9x}.  After  this  process,
VOSA identified 377 WD candidates  to show IR-excess among our initial
3,733 objects.

In a first step, we  visually inspected the optical (Pan-STARRS1, SDSS
and  DSS) and  IR (2MASS  and WISE)  images of  the 377  sources using
Aladin\footnote{http://aladin.u-strasbg.fr} \cite[][]{Bonnarel00}. The
VOSA SED fittings were also checked. We removed a total of 221 sources
(58\%)  from our  target  list mainly  due to  the  WISE poor  spatial
resolution (6\arcsec\,  beam size), which causes  a significant number
of  false  positives due  to  contamination  by nearby  sources.  This
contamination rate  is slightly lower  to that found by  other authors
\citep[75\%, e.g.][]{Barber16}. Of the  remaining 156 objects, 38 were
identified as  co-moving systems  by using  {\it Gaia}  parallaxes and
proper  motions,  as  well  as   photometry  from  surveys  where  the
components of  the systems  appear spatially  resolved. Most  of these
objects  can be  easily identified  by the  jump in  their SEDs  at IR
(mainly  WISE) wavelengths.  This jump  is caused  by the  sum of  the
fluxes of  the nearby  sources that  form the system  due to  the WISE
spatial resolution  (see an  example in Figure\,\ref{fig:bin}).  26 of
the 38 co-moving  pairs were identified as WD+M systems,  most of them
already        reported        by        \citet{Badry2018}        (see
Table\,\ref{t:co-moving}). Effective temperatures were estimated using
VOSA and  the \citet{Koester10}  and BT-Settl  \citep{Allard12} models
for  the WD  and the  M star  components, respectively.  The other  12
systems are reported in Table\,\ref{t:cmovingknown}.

Of the remaining  118 WDs, one has associated two  entries in the {\it
  Gaia} DR2  catalogue separated  by less  than 2\arcsec\,  (Gaia IDs:
883243467325018496   /   883243467323599616).   The   differences   in
parallaxes  and the  SED analysis  made  with VOSA  conclude that  the
secondary  component  causing  the  IR   excess  is,  most  likely,  a
background M giant. Similar cases are the sources 2612592841965015424,
2564424130905288192,      63846445499673472,      2969841487138850560,
5657351404992422784, 3650552739370519680  and 1316268323580640256, the
latter studied by \citet{Barber14} who  confirmed the IR excess arises
due to the contamination of a background object.

From the final list of 110  selected WD candidates, 77 benefit from IR
photometry at both near  and mid IR wavelengths and 33  just at mid IR
wavelengths.   Tables\,\ref{t:dusty}-\ref{t:dusty2}  list  the  77  WD
selected candidates with  available near and mid IR  photometry, 52 of
which are new discoveries not yet  known to host disks, brown dwarf or
low-mass  companions. The  effective temperatures  and radii  (derived
from   the   Stefan-Boltzmann   equation)   associated   to   the   IR
contribution's   sources  are   estimated  from   the  composite   SED
fitting. For this, VOSA uses a  range of values around the white dwarf
\teff$ $  and $\log{g}$ obtained  from the  single best fit  using the
\citet{Koester10}  models together  with  a blackbody  with \teff  $<$
5,000\,K. For WDs,  the errors in effective  temperatures arising from
the composite SED fitting are given by the step of the grid of models,
which changes with \teff. The step  in the grid of blackbody models is
set to 25\,K.

Table\,\ref{t:dusty3}  summarizes the  information of  the 33  targets
having only  IR photometry at  mid IR  wavelengths. The SEDs  of these
IR-excess candidates are poorly populated and hence it is difficult to
asses whether or not the detected excesses are real. Another important
issue of only having  at hand mid IR photometry is that  it is hard to
asses if  the excesses arise  from a dust disk  or a companion.  It is
worth noting that  for one of them ({\it  Gaia} ID 128198912443928691)
the IR excess  has been confirmed to arise due  to a circumbinary disk
\citep{Farihi08}.  However,  to  avoid  including a  large  number  of
potential false positive detections in our list, these WDs will not be
considered further in this work.

More information about the final 77 selected candidates for displaying
IR  excess  with  available  photometry   at  both  near  and  mid  IR
walelengths, including  a visualization of  their SED fitting,  can be
found  at the  SVO  archive of  WDs  with IR  excess  from {\it  Gaia}
DR2\footnote{http://svo2.cab.inta-csic.es/vocats/v2/wdw3}.

\begin{table}
\centering
\caption{List  of 12  binary/multiple  co-moving pairs  found in  this
  work.}
\label{t:cmovingknown}
\setlength{\tabcolsep}{0.8ex}
\begin{tabular}{ccc}
\hline
Identifier & Source & Comment \\
\hline
eps Ret   & \cite{Farihi2011} & (1)  \\
eps Ret B & &   \\
eps Ret b & &   \\
PM\,J04032+2520 & \cite{Limoges2015} & (2) \\
PM\,J04032+2520E & & \\
2MASS\,J04031652+2520192 & & \\
EGGR\,576 & \cite{Gianninas2011} & WD+WD \\
EGGR\,577 & & \\
LP\,402-28 & Simbad & WD+WD \\
LP\,402-29 & & \\
SDSS\,J230249.37+243027.9 & Simbad & (3) \\
SDSS\,J230250.37+243013.3 & & \\
L\,462-56A & Simbad & WD+WD \\
L\,462-56B & & \\
Gaia\,2751252493861856000 & \cite{Badry2018} & WD+WD (4)\\
Gaia\,2751252489566343680 & & \\
Gaia\,3404213863611804672 & \cite{Badry2018} & WD+WD (4)\\
Gaia\,3404213863614488192 & & \\
 Gaia\,4209104513139995136 & this work & WD+WD (5) \\
 Gaia\,4209104577563403136 & & \\
 Gaia\,4659809928696442368 & this work & WD+WD (5) \\
 Gaia\,4659809928696442496 & & \\
Gaia\,4964509614631078400 & \cite{Badry2018} & WD+WD (5)\\
Gaia\,4964509614631078272 & & \\
 Gaia\,5184384997855024384 & this work & WD+WD (5,6) \\
 Gaia\,5184385002150373632 & & \\
\hline
\end{tabular}
\begin{list}{}{} 
\item[]
(1): Binary system formed by a red giant (K2\,III) with a confirmed extra-solar planet and a white dwarf (DA3).
\item[]
(2) Triple system formed by two white dwarfs and a M-dwarf.
\item[]
(3) WD+high proper motion object.
\item[]
(4): Both components included in \cite{Jimenez-Esteban18}.
\item[]
(5): First component included in \cite{Jimenez-Esteban18}. WD nature of the second component derived from its position in a M$_{G}$ vs $G$-$G_\mathrm{RP}$ diagram.
\item[]
(6): With a $G$-$G_\mathrm{RP}$ = 0.92, the second component is one of the coolest WDs in our sample.
\end{list}
\end{table}

\begin{table*}
\centering
\caption{List of  the 77 WDs found  in this work displaying  IR excess
  and with at  least three photometric points spread  out between near
  and mid IR  wavelengths. 52 are new  discovery candidates. Effective
  temperatures were calculated as described in Sect.\,\ref{Iden} while
  coordinates have  been taken  from the SVO  archive of  White Dwarfs
  from  {\it  Gaia}. Surface  gravities  and  masses are  obtained  as
  described in Sect.\,\ref{s-disc}.  Note the blackbody radii provided
  for the dusty  WDs have no physical meaning. For  WDs, the errors in
  effective temperatures are given by the  step of the grid of models,
  which changes with  \teff. The step in the grid  of blackbody models
  is  set to  25\,K.  The  second-last column  indicates the  expected
  cause of the  IR excess (either circumstellar disk  or companion; in
  italics) resulting from  the visual SED inspection.  The last column
  indicates    the    same   but    based    on    the   IR    colours
  (Fig.\,\ref{fig:excess}) and  the mass of  the WD (in  italics; note
  the classification  based on the WD  mass is only provided  for four
  low-mass WDs  expected to be in  close binaries and it  is indicated
  after the IR colour classification  following /). Also indicated are
  the confirmed  disks and  brown dwarfs by  other studies.   In these
  cases the classifications are not given  in italics. In bold face we
  indicate the assumed final classification for each object.}
\label{t:dusty}
\setlength{\tabcolsep}{0.7ex}
\begin{tabular}{cccccccccccc} 
\hline
Gaia ID  & RA      & DEC   & Name & \teff& \teff & R  & logg & Mass & Ref. & Type & Type\\
         & (ICRS)  &(ICRS) &      & (WD)& (bb)  & (bb) & (WD) & (WD)&      &      &\\
DR2 &  deg & deg & & K & K & R$_{\odot}$ & dex & \Msun & & SED & other\\
\hline
2416481783371550976 & 1.8951 & -16.0921 & EGGR 509 & 12000$\pm125$ & 2450 & 0.07& 7.84 &0.52 & & \emph{comp.} & \emph{disk}\\
2798132572998105984 & 1.9484 & 19.8568& & 12250$\pm$125 & 2500&0.02 & 7.75 & 0.48 & & {\bf \emph{comp.}} & \emph{?}\\
367949367212923392 & 12.3001 & 38.6918 & LAMOST J004912.04+384129.8 & 9500$\pm$125 & 850 & 0.23&7.76 &0.48 & &\emph{?} &\\
2529337507976700928 & 12.6909 & -3.4487 & & 20000$\pm$310 & 1050 & 0.10& 8.92&1.20 & &{\bf \emph{disk}} &\\
  5026963661794939520 & 17.1503 & -32.6288 & HE 0106-3253 & 15750$\pm$125&1900 & 0.06& 7.97 & 0.60  & (Fa10) & \emph{disk} & {\bf disk}\\
  2354670057156360576 & 17.3882 & -19.0215 & & 14250$\pm$125 & 550 & 0.48& 7.93 &0.57 & (De17)& \emph{?} & \\
  4913589203924379776 & 18.0888 & -56.2411 & JL 234                     & 18250$\pm$125&950 & 0.18& 7.89 & 0.56 & (Gi12) & \emph{disk} &{\bf disk}\\
  2593884960855727872 & 21.2525 & 18.1945 & & 8750$\pm$125 & 1700 & 0.05&8.32 &0.79 & &{\bf \emph{disk}}& \emph{disk}\\
  2588874825669925504 & 23.8868 & 14.7649& LSPM J0135+1445 & 8250$\pm$125&2450& 0.06& 7.50 & 0.37 & (St13) & \emph{comp.} &{\bf BD}/\emph{comp.}\\
  5135466183642594304 & 26.8412 & -21.9477 & GD 1400                    & 12000$\pm$125&2500 & 0.06& 8.11 & 0.67 & (Fa04)& \emph{comp.} &{\bf BD}\\ 
  291057843317534464(*)& 26.9784 & 23.6617  & WD 0145+234          & 12500$\pm$125&-& -& 7.99 & 0.60 & &{\bf \emph{disk}}& \emph{disk}\\
  95297185335797120   & 27.2377 & 19.0405  & Wolf 88                    & 13250$\pm$125&2200 & 0.06& 8.29 & 0.78 & (Fa09) & \emph{disk} &{\bf disk} \\ 
  4632284754595134080 & 31.3539 & -79.6844 & & 9500$\pm$125 & 800 &0.14 &8.09 &0.65 & &{\bf \emph{disk}}& \emph{disk}\\
 2489533370280291584 & 35.8356 &  -4.9852 & & 10250$\pm$125 & 600 & 0.45& 8.05& 0.63& &{\bf \emph{disk}}&\\
 2489275328645218560 & 38.5646 & -4.1026 & & 13500$\pm$125 & 900 & 0.16& 8.12 &0.67 & &{\bf \emph{disk}}& \emph{disk}\\
  5187830356195791488 & 45.7209 & -1.1429  & GD 40                      & 14500$\pm$125&950 & 0.16& 8.16& 0.70& (Ju07)& \emph{disk}& {\bf disk}\\  
  139331247344776832 & 47.0822 & 36.4914 & & 7500$\pm$125 & 1650 & 0.01& 7.94 & 0.56& &\emph{?}&\\
  4833891614684676736 & 52.3630 & -47.6435 & & 9750$\pm$125 & 1450 & 0.06& 7.84 &0.52 & &{\bf \emph{disk}}& \emph{disk}\\
 542865797290276352 & 54.1894 &  70.7364 & & 10500$\pm$125 & 2800 &0.03 & 7.99 &0.60 & &\emph{?}&\\
    3251748915515143296 & 62.7590 & -3.9735  & GD 56                      & 14500$\pm$125&950 & 0.45& 8.00 & 0.61 & (Ju07)& \emph{disk}& {\bf disk}\\
  4837423353408638080 & 63.2121 & -45.1696 &                            & 14000$\pm$125&1350 & 0.08& 8.02 & 0.62 & & {\bf \emph{disk}}&\emph{disk}\\
  4653404070862114176 & 64.9077 & -73.0623 & [DI91] 1592                & 18250$\pm$125&1350 & 0.02& 7.92 & 0.58 & (Ho13)&  {\bf \emph{disk}} &\\
  271992414775824640  & 66.0653 & 52.1696  & KPD 0420+5203              & 15250$\pm$125&1100& 0.18& 8.09 & 0.67 & (Ba16)& \emph{disk} & {\bf disk}\\
  152740654933891072 & 68.4777 & 28.4579 & PM J04339+2827 & 14000$\pm$125&1850 & 0.02& 7.92 & 0.57&(Xu15) & \emph{comp.} & {\bf \emph{BD}}\\
  203931163247581184 & 69.6641 & 41.1585 & GD 61 & 14250$\pm$125&1350& 0.001 & 8.08 & 0.66 & (Fa 11) & \emph{disk} &{\bf disk}\\
  2986304298645920384 & 75.3165 & -15.1900 & & 11250$\pm$125 & 2350 & 0.05&7.83 &0.51 & &\emph{?}&\\
  3415788525598117248 & 77.5087 & 23.2613  & LAMOST J051002.11+231541.0 & 17250$\pm$125&1200 & 0.13& 8.11 & 0.68 & &{\bf \emph{disk}}& \emph{disk}\\ 
  4799224635833122304 & 82.7521 & -45.9670 & & 10750$\pm$125 & 850 &0.13 &8.02 &0.61 & &{\bf \emph{disk}}&\\
  4795556287084999552  & 84.4726 & -47.9681 & EC 05365-4759              & 18250$\pm$125&450 & 1.40& 7.87& 0.55& (De16)& \emph{disk} &{\bf disk} \\ 
      3329569015639064192(*) & 90.6529 & 9.07322 & LSPM J0602+0904 & 6000$\pm$125 & - & -& 7.50& 0.35& &\emph{disk} &\emph{?/comp.}\\
    962995581174346112  & 90.7863 & 45.3077  &                            & 14750$\pm$125 &1950& 0.04& 7.95 & 0.58 & &{\bf \emph{disk}}& \emph{disk}\\ 
    3112786176370258688 & 105.6910& 0.0552   &                      & 11750$\pm$125&750 & 0.14& 8.03& 0.62& &{\bf \emph{disk}}&\emph{disk}\\
    5490140356700680576 & 108.6243 & -55.6572& & 9000$\pm$125 & 2700 & 0.07& 7.64& 0.43& &\emph{comp.}& {\bf \emph{BD}}/\emph{comp.}\\
  872009447786700672(*) &  112.5000 & 27.2781 & LSPM J0730+2716W & 9250$\pm$125 & - & -&7.98 &0.59 & &{\bf \emph{disk}}& \emph{disk}\\
  5292685793681027968 &  113.6708 & -60.1979 & & 9250$\pm$125 & 800 & 0.12&8.24 &0.74 & &{\bf \emph{disk}}&\emph{disk}\\
  1081504483467714176      & 120.6156 & 56.5321 & & 10750$\pm$125 & 950 & 0.16 & 8.06&0.64 & &{\bf \emph{disk}}& \emph{disk}\\
585513959248023936 & 141.1386 & 5.3519 & & 6000$\pm$125 & 950 & 0.09&8.18 &0.70 & &{\bf \emph{disk}}&\\
5662556012001458944 & 141.2082 &  -24.38456 & 0.23& 8500$\pm$125 & 700 & 0.24&7.96 &0.57 & &{\bf \emph{disk}}&\emph{disk}\\
 5740372469987778304 & 143.4221 & -10.0026 & & 8000$\pm$125 & 1450 &0.06 & 8.56& 0.95& &\emph{?}&\\
5459131788043369344 & 154.3688 &  -32.6025 & & 8000$\pm$125 & 1300 &0.05 & 7.93 &0.56 & &{\bf \emph{disk}}& \emph{disk}\\
    3888723386196630784 & 154.5154 & 15.8660  & PG 1015+161               & 21000$\pm$500&1600& 0.01& 7.98 & 0.61 & (Ju07) &\emph{disk}& {\bf disk}\\ 
    3810933247769901696 & 169.8012 & 2.3426   & GD 133                    & 12250$\pm$125&900& 0.17& 8.01&0.61 & (Ho13) &\emph{disk} & {\bf disk}\\ 
  771517005584473600 & 171.424 & 42.3930 & GD 308 & 9500$\pm$125 & 1400 & 0.06& 8.14&0.68  &(De11) &{\bf \emph{disk}}& \emph{disk}\\
   3571559292842744960 & 178.3134 & -15.6104& EC 11507-1519 & 11000$\pm$125 &750& 0.53& 7.90 & 0.55 & (Ho13) & \emph{disk} &{\bf disk}\\
  3543074313820703488 & 178.5127 & -19.2376 & & 7250$\pm$125 & 750 & 0.28 &7.77 &0.48  & &{\bf \emph{disk}}&\\
 3479615106870788864(*) &  178.5145 & -31.0292 & & 6250$\pm$125 & -& - & 8.17& 0.69& &{\bf \emph{disk}}& \emph{disk}\\
 4028120776036373760 & 180.4780 & 34.0154 &  SDSS J120154.70+340055.9 & 6000$\pm$125 & 1300 & 0.06& 8.14& 0.67& &\emph{?}&\\
 1692520339315508224 & 184.2221 & 74.9237 & & 6500$\pm$125 & 1250 & 0.05& 8.00&0.59 & &{\bf \emph{disk}}& \emph{disk}\\
 3903151246497510784 &  190.1509 & 9.5361 & & 6750$\pm$125 & 750 & 0.14& 8.45& 0.87& &{\bf \emph{disk}}&\\
  3663900436870097664 & 208.7495 & 1.1387 & SDSS J135459.89+010819.3 & 11500$\pm$125 & 750 & 0.20&7.88 &0.54 & & {\bf \emph{disk}}&\emph{disk}\\
 1494157691363079168(*) & 217.1406 & 44.0630 & & 7500$\pm$125 & -& -& 8.36& 0.81& (De11) &{\bf \emph{disk}}&\emph{disk}\\
 1488904946359359488 & 222.5277 & 40.9264 & CBS204 & 13500$\pm$125 & 800 & 0.21&7.88 &0.54 & &\emph{?}&\\
   1183473535423719296     & 227.4253 & 14.1892 & & 6250$\pm$125 & 1250 & 0.04 &8.15 &0.67 &  &{\bf \emph{disk}}& \emph{disk}\\
   6315417253178248960 &  229.6198 & -11.8109 & & 10250$\pm$125 & 700 & 0.23& 7.86 &0.52 & &{\bf \emph{disk}}&\\
\hline
\end{tabular}
\end{table*}

\begin{table*}
\centering
\caption{List of  the 77 WDs found  in this work displaying  IR excess
  and with at  least three photometric points spread  out between near
  and mid IR wavelengths (cont.).}
\label{t:dusty2}
\setlength{\tabcolsep}{0.7ex}
\begin{tabular}{cccccccccccc} 
\hline
Gaia ID  & RA      & DEC   & Name & \teff& \teff & R  & logg & Mass & Ref. & Type & Type\\
         & (ICRS)  &(ICRS) &      & (WD)& (bb)  & (bb) & (WD) & (WD)&      &      &\\
DR2 &  deg & deg & & K & K & R$_{\odot}$ & dex & \Msun & & SED & other\\
\hline
   1641326979142898048 & 235.4372 & 64.8978 & V* KX Dra & 11000$\pm$125&850 & 0.27 & 7.90 & 0.55& (Ki12) & \emph{disk} &{\bf disk}\\
  1429618420396285952 & 243.3191 & 55.3572  & SBSS 1612+554             & 11250$\pm$125&950& 0.16& 8.06&0.64 & &{\bf \emph{disk}}& \emph{disk}\\
4390134326651497728 & 260.957 & 4.9799 & PM J17238+0458 & 8500$\pm$125 & 550 & 0.35&7.94 &0.56 & &{\bf \emph{disk}}& \emph{disk}\\
 1368236912466084352(*) &265.7298 &  51.6024& SDSS J174255.14+513608.4 & 8750$\pm$125 & - & - &7.95 &0.57 &  &{\bf \emph{disk}}& \emph{disk}\\
  4583221109793391232(*) & 271.6897 & 27.5299 & & 6500$\pm$125 & - &- & 8.09 &0.64 & & {\bf \emph{disk}}& \emph{disk}\\
  6417955993895552128 & 273.5734 & -73.9174 &  & 7750$\pm$125 & 700 &0.80 &  8.06 & 0.63 & &{\bf \emph{disk}}& \emph{?}\\
 2155960371551164416 & 285.8315 &  60.5980 & GD532 & 10750$\pm$125 & 850 & 0.23&8.04 &0.63 & &{\bf \emph{disk}}&\\
   6429048245152936320 & 305.0718& -65.4240 & & 6500$\pm$125 & 1900 & 0.05& 8.25 & 0.74 & &{\bf \emph{comp.}}& \emph{?}\\   
    1837948790953103232 & 315.1447 & 21.3826  &                           & 15250$\pm$125&1000 & 0.44& 7.91 & 0.59 & &{\bf \emph{disk}}& \emph{?}\\  
  6462911897617050240 & 319.9055 & -55.8382 & LAWD 84 & 9500$\pm$125&650 & 0.10& 8.02 & 0.61 & (Fa09) & \emph{disk}& {\bf disk}\\
  6580498481454705408 & 320.3473 & -42.1484 & & 7500$\pm$125 & 600 & 0.26& 8.32& 0.79& &{\bf \emph{disk}}&\\
  6811977801160882944 &  328.3816 & -26.4821 & & 14750$\pm$125 & 700 & 0.23&9.09 &1.31 & &{\bf \emph{disk}}&\\
  2677851743291189888 & 335.1279 & -0.6854 & PHL 5038 & 7500$\pm$125 & 1050 & 0.12 & 7.87& 0.53& (De11) &{\bf \emph{disk}}& \emph{disk}\\
  2595728287804350720 & 336.0726 & -16.2631 & PHL 5103 & 10000$\pm$125 & 1450 & 0.04&8.12 &0.67 & (Ro15)& \emph{disk} &{\bf disk}\\
   1900545847646195840 & 337.4920 & 30.4028 & PM J22299+3024 & 10500$\pm$125 & 2550 & 0.08&7.41 &0.35 & &{\bf \emph{comp.}}& \emph{disk/comp.}\\
   2622979271185741312 & 338.3480 & -6.0278 & & 8250$\pm$125 & 850 & 0.31&8.07 &0.63 & &{\bf \emph{disk}}&\\
  6594180460552162944 &  338.477 & -38.5436 & LP 1033-28 & 8500$\pm$125 & 1000 & 0.07& 8.05& 0.62& &{\bf \emph{disk}}&\emph{disk}\\
     2712240064671438720 & 344.3588 & 7.9285 &  G28-27 & 13750$\pm$125 & 950 & 0.07& 9.36& 1.44& (De11) &{\bf \emph{disk}}&\\
  1995097319287822080 & 346.3820 & 51.4227 & & 12750$\pm$125 & 1800 & 0.04&7.93 &0.57 & &{\bf \emph{disk}}&\emph{disk}\\
 6499095244738784128 &  349.0564 &  -55.4912 & & 10250$\pm$125 & 950& 0.10& 8.35&0.81 & &{\bf \emph{disk}}&\emph{disk}\\
   2660358032257156736 & 352.1975 & 5.2478 & V* ZZ Psc & 10750$\pm$125&950 & 0.19 & 7.90 & 0.55 & (Re05)& \emph{disk} &{\bf disk}\\   
  1923682286712356992 & 352.9001 & 41.0248  & EGGR 160                  & 14500$\pm$125&700 & 0.36& 7.94& 0.58& (Ho13)& \emph{disk} &{\bf disk}\\
   6538863343364422528 & 355.1527 & -37.1458 & EC 23379-3725 & 12250$\pm$125 & 700 & 0.41& 7.74& 0.48& &{\bf \emph{disk}}& \emph{disk}\\
\hline
\end{tabular}
\begin{list}{}{}
\item[Comments:]
(*) Marginal IR excess. No reliable values of blackbody effective temperature and radius.
\item[References:]
(Fa10): \cite{Farihi10},
(Gi12): \cite{Girven12},
(St13): \cite {Steele13}
(Fa04):  \cite{Farihi2004}
(Fa09): \cite{Farihi09},
(Ju07): \cite{Jura07},
(Ho13): \cite{Hoard13},
(Ba16): \cite{Barber16},
(Xu15): \cite{Xu2015}
(Fa11): \cite{Farihi2011},
(De16): \cite{Dennihy16},
(Fa08): \cite{Farihi08b},
(Ki12) \cite{Kilic12},
(De11): \cite{Debes2011},
(De17): \cite{Dennihy17}
(Re05): \cite{Reach05}.
(Ro15): \cite{Rocchetto15}.
\end{list}
\end{table*}

\begin{table}
\centering
\caption{List of 33 IR-excess  WD candidates with available photometry
  at only mid IR wavelengths.}
\label{t:dusty3}
\setlength{\tabcolsep}{0.3ex}
\begin{tabular}{cccc} 
\hline
Gaia ID  & RA  & DEC  & Name \\
         &(ICRS) & (ICRS) & \\
DR2 &  deg & deg &  \\
\hline
2741440172922171008 & 4.2305 & 5.0786 & SDSS J001655.37+050442.1 \\
2858896086675180928 & 7.2235 & 30.0918 & \\
2344098385998773120 & 11.3746 & -25.0516 & \\
  377520826387065856  & 13.0183 & 45.0927  &        \\
  306805388153226368 & 16.7916 & 27.1691 & \\
  306350606950880128 & 16.8592 & 25.3099 & \\
 2790417540424293120 & 16.9558 & 21.1294 & SDSS J010749.34+210745.2 \\
 5161531373793965440 & 49.6922 &  -13.0005 & \\
  568168544844912128 & 57.0280 & 80.8102 & \\
   4887631143142117632 & 57.4624 & -30.5289 & \\
    498487545190443136 & 89.8444 & 72.9873 & \\
 992771180686912000 & 98.8498 &  52.2593 & \\
   921804126089222784 & 122.9556 & 42.2025 & KUV 08084+4221 \\
   1118374024628715264 & 129.4702 & 69.2181 & \\
  1051954485699665280$^{1}$ & 152.5329 & 61.9211 & \\
  738060065046666240 & 163.0523 & 33.3884 & \\
   789712823515276416 &  169.3462 & 48.8665 & \\
  3705386281897262848 & 193.0632 & 4.1786 & HS 1249+0426 \\
  3938156295111047680 & 196.4251 & 18.0179  & V* GP Com                 \\
   1281989124439286912$^{2}$ & 224.5277 & 29.6223 &EGGR 298 \\
  1157317008497672320 & 227.7374 & 6.4638 & SDSS J151056.99+062749.7 \\
1219699145026398848     & 237.2293 & 24.8536 & SDSS J154855.04+245112.9 \\
  1316607896578157824$^{3}$ & 240.4179 & 27.5969 & LSPM J1601+2735 \\
 1199686173677816576$^{1}$ &  242.1648 & 17.3935 & \\
1428562506980546688 & 244.1432 & 54.1698 & \\
4555079659441944960 & 262.1905 & 20.8949 & \\
 6845706900891884928 & 303.9679 & -28.5888 & \\
   6886271973655421824 & 312.8046 & -15.3495 & \\
  6809396800693752576 &  314.4396 & -20.0568 & \\
   6884996230921934976 & 316.6775 &-14.7588 & \\
   2284456545980836736 & 319.6924 & 76.9831 & \\
  1946495125767488896 & 326.3801 & 32.8618 & \\
   2199338643594260352    & 329.6359 & 58.0752 & Lan 432 \\
\hline
\end{tabular}
\begin{list}{}{}
\item[References:]
(2): \cite{Farihi08b},
(1): \cite{Debes2011},
(3): \cite{Dennihy17}
\end{list}
\end{table}

\section{Methodology assessment}
\label{method}

The  efficiency  of  our  methodology was  assessed  using  the  false
negative rate, i.e. the fraction of  known IR-excess WDs that were not
rediscovered in  our search. In particular,  we compiled a list  of 24
WDs at less than  100 pc and with IR excess  confirmed by Spitzer.  20
objects (83\%) were identified using  our methodology and are included
in  Tables\,\ref{t:dusty}-\ref{t:dusty3}. The  remaining four  objects
were not identified due to the following two reasons:

(1) Contamination  of the  WISE photometry  due to  the presence  of a
nearby source  (WD1929+011 and  WD0950-572). It  is worth  noting that
these  two objects  are not  included in  the SEIP  Spitzer catalogue,
otherwise VOSA would have very likely detected the IR excess.

(2)  Unreliable WISE  photometry in  W3 and  W4 bands  (WD2132+096 and
WD2328+107).  These  targets are  not  included  in the  SEIP  Spitzer
catalogue neither, hence VOSA could not detect the IR excess.

These results  confirm the  robustness of  our methodology  (a success
rate of 83\%) to identify WDs  with IR excess. We therefore assume our
IR-excess sample  to be 83\% complete.   It is also important  to note
that if we take  into account that the four objects  that we could not
recover were  identified from publicly unavailable  Spitzer data (i.e.
data not included in the SEIP catalogue), then the success rate of our
method would increase to 100\%.

Assuming an  IR-excess completeness  of 83\% for  our sample  does not
imply that  we have discovered  83\% of  all WDs displaying  IR excess
within 100  pc from  the Sun.   First, we  are basing  our methodology
assessment  on  \emph{confirmed}  WDs   with  IR  excess  by  Spitzer,
therefore there may  exist WDs without Spitzer data  and displaying IR
excess that we  are not taking into account and  that are consequently
missed by this and all  previous studies.  Second, and more important,
we are only considering 3,733  objects with reliable IR photometry and
with $G_{\rm  BP}-G_{\rm RP}$  colours below 0.8  mag within  our {\it
  Gaia} 100 pc sample.

\section{Characterization of the sample}
\label{charac}

It  is widely  accepted that  there  are three  main possibilities  to
explain  the IR  excess  detections in  WDs such  as  those listed  in
Tables\,\ref{t:dusty}-\ref{t:dusty2}:  the presence  of a  brown dwarf
companion,  the  presence  of   low-mass  stellar  companion  and  the
existence of a  circumstellar dust disk. To discern the  origin of the
IR excess  we follow  two different  approaches, namely  the use  of a
colour-colour  diagram and  the  visual inspection  of  the SEDs.  The
classification  of IR-excess  origin based  on these  two methods  are
provided    in     the    last    and    second-last     columns    of
Tables\,\ref{t:dusty}-\ref{t:dusty2}, respectively.

\begin{figure}
        \includegraphics[width=\columnwidth]{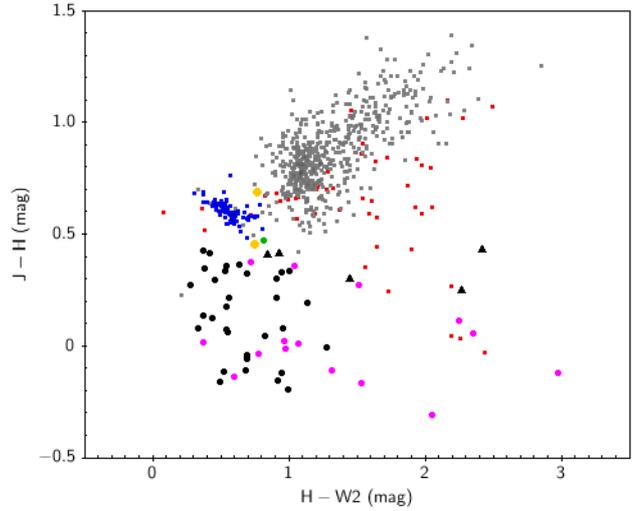}
    \caption{$J$-$H$  vs $H$-$W2$  colour-colour  diagram showing  the
      loci of  stars and  brown dwarfs  of M  (blue squares),  L (grey
      squares) and T (red squares)  spectral types. M dwarfs have been
      taken  from   \citet{Reiners18}  while  L,T  objects   are  from
      \citet{Smart17}.  Bullets  represent the  54 WDs with  IR excess
      listed in  Tables\,\ref{t:dusty}-\ref{t:dusty2} with  good 2MASS
      and  WISE  photometry.   Pink  bullets  (15)  represent  objects
      classified  as  confirmed  dusty  WDs  in  the  last  column  of
      Tables\,\ref{t:dusty}-\ref{t:dusty2},  while  green bullets  (1)
      are objects  classified as confirmed  WD+BD systems in  the same
      column. Black (31) and yellow (2) bullets are objects classified
      as dusty WDs  or WD+BD systems, respectively  according to their
      position in the diagram.   Black triangles (5) represent objects
      for which the origin of the IR excess is uncertain.}
    \label{fig:excess}   
    \end{figure}

We  use  the  colour-colour   diagram  proposed  by  \citet{Barber14}.
Fig.\ref{fig:excess} compares the IR colours of the 54 WDs with excess
included in  Tables\,\ref{t:dusty}-\ref{t:dusty2} and with  good 2MASS
(Qflg $\neq$ U  in the J and  H bands) and WISE (ccf=0  and qph=A/B in
the W2  band) photometry  to those of  M, L, T  dwarf stars  and brown
dwarfs.  15  objects (pink  bullets in  Figure\,\ref{fig:excess}) have
been classified as  dusty WDs in the literature, while  for one object
(green bullet) the IR excess has been ascribed to the presence of a BD
companion. 31 objects (black bullets)  are located well apart from the
stellar/brown  dwarf loci,  and we  hence classify  them as  dusty WDs
based on  their IR colours.   Two objects (yellow bullets)  lie either
near the brown dwarf locus or near a confirmed WD+BD binary as we thus
classify them as WD+BD candidates. For  one of these objects (Gaia ID:
152740654933891072)  the  morphology  flag  of  the  UKIDSS  catalogue
indicates the possibility  that this source was extended  at more than
one  bandpass,   which  is  an  additional   evidence  supporting  the
hypothesis  that the  IR  excess is  due to  nearby  brown dwarf.  The
remaining five (black triangles) lie  in colour regions expected for T
brown dwarfs where also confirmed  dusty WDs are located, therefore it
becomes  difficult to  asses  the origin  of the  IR  excess in  these
objects.  We explored their quality flags associated to the {\it Gaia}
astrometry.   Following the  latest recommendations  published by  the
{\it Gaia} ESA team  in the {\it Known issues with  the {\it Gaia} DR2
  data}                       web                       page\footnote{
  https://www.cosmos.esa.int/web/gaia/dr2-known-issues},   we  defined
sources with  good astrometry as  those having the  re-normalised unit
weight error (RUWE) $<$ 1.4. Sources having a higher value of RUWE may
have a  worse astrometric  solution due to  different effects,  one of
those being the presence of a close companion. However, we got a value
of  RUWE$<$1.4  for   the  five  objects,  which   prevented  us  from
discriminating between a  disk or brown dwarf/low-mass  star origin of
their IR excesses.

In order to further explore the  origin of the IR excesses we visually
inspected the  SEDs of the 77  candidates. Eight of them  benefit only
from  J  2MASS and  WISE  photometry,  hence  we  do not  have  enough
information at hand to reach a  conclusion. In these cases we flag the
origin of the excess as unknown (\emph{?} in the second last column of
Tables\,\ref{t:dusty}-\ref{t:dusty2}).  For  eight additional  objects
the IR excesses arise at the J and/or H bands, which supports the idea
of these  WDs being in  binary systems  with either low-mass  stars or
brown    dwarf    companions.    Two    objects    (\emph{Gaia}    IDs
2588874825669925504  and 5135466183642594304)  are in  fact confirmed,
and one  more (152740654933891072) is  a candidate to harbour  a brown
dwarf based on its  IR colours (Fig.\,\ref{fig:excess}). These objects
are  flagged  to  harbour  companions in  the  second-last  column  of
Tables\,\ref{t:dusty}-\ref{t:dusty2}  and we  include the  brown dwarf
classification of  the three discussed  targets in the last  column of
the same  tables. The SEDs  of 61 WDs display  IR excess at  K-band or
longer  wavelengths,  characteristic  features of  circumstellar  disk
candidates. These objects are hence flagged as such in the second-last
column of Tables\,\ref{t:dusty}-\ref{t:dusty2}.

\section{Discussion}
\label{s-disc}

We have identified 77 IR-excess  WD candidates within the 100\,pc {\it
  Gaia} WD catalogue, 52 of which  have not been published before (see
Tables\,\ref{t:dusty}--\ref{t:dusty2}). The  WD effective temperatures
are  derived  fitting  the  photometric  SED of  each  WD  using  VOSA
(Section\,\ref{Iden}), which together with the bolometric luminosities
(also provided  by VOSA by  making use  of the {\it  Gaia} parallaxes)
yield the radii of the WDs  from the Stefan-Boltzmann equation. The WD
masses   are  estimated   interpolating   the   radii  and   effective
temperatures  obtained  in this  way  in  the  WD cooling  tracks  for
hydrogen-rich,  DA, atmospheres  of \citet{Renedo2010},  following our
procedure  described  in  \citet{Jimenez-Esteban18}. Given  that  {\it
  Gaia} DR2 does  not provide stellar spectra, the  reliability of the
determined  masses relies  on  the  assumption that  all  our WDs  are
DAs. As we showed in \citet{Jimenez-Esteban18}, the contribution of DB
(helium-rich)  WDs  to  the  sample of  WDs  with  determined  stellar
parameters  should  be  no   higher  than  $\simeq$6\%.   The  average
uncertainties  that  are  introduced  when  assuming  DA  models  when
deriving  the  effective  temperatures  and   masses  of  DB  WDs  are
1700$\pm$1680\,K  and  0.08$\pm$0.06\Msun,   respectively,  where  the
errors  are  the  standard  deviations.   These  values  are  obtained
comparing      the      stellar     parameters      calculated      by
\citet{Jimenez-Esteban18}  for   152  common   SDSS  DB  WDs   in  the
spectroscopic  catalogue of  \citet{Koester2015}  with relative  error
below 10\% in  effective temperature and surface  gravity errors below
0.05 dex.

Four WDs have determined  masses under 0.45\,M$_\mathrm{\odot}$, which
indicates  these objects  are likely  members of  close binaries  with
sub-stellar or  low-mass companions (from  which the IR  excess likely
arises)  that  formed  through   common  envelope  evolution.  Another
possibility is  that these are  non-DA WDs,  in which case  the masses
cannot be  considered as reliable.  The four potentially  low-mass WDs
have    the   following    {\it    Gaia}   IDs:    2588874825669925504
(LSPM\,J0135+1445),     3329569015639064192     (LSPM     J0602+0904),
5490140356700680576,              1900545847646195840              (PM
J22299+3024).   LSPM\,J0135+1445   is   a   confirmed   WD+BD   binary
\citep{Steele13}.   3329569015639064192   is  classified  as   a  disk
candidate based on  its SED, which supports the idea  of this WD being
of a  non-DA spectral  class. 5490140356700680576  is classified  as a
companion and  brown dwarf  by its SED  and IR  colours, respectively,
which  strongly  indicates   this  WD  is  indeed  part   of  a  close
binary.  Finally,  the  SED  and  IR  colours  of  1900545847646195840
indicate a companion and disk origin, respectively, and therefore more
information is required  (e.g. IR spectroscopy) to confirm  this WD is
in a close binary system.

\begin{figure}
    \includegraphics[width=\columnwidth]{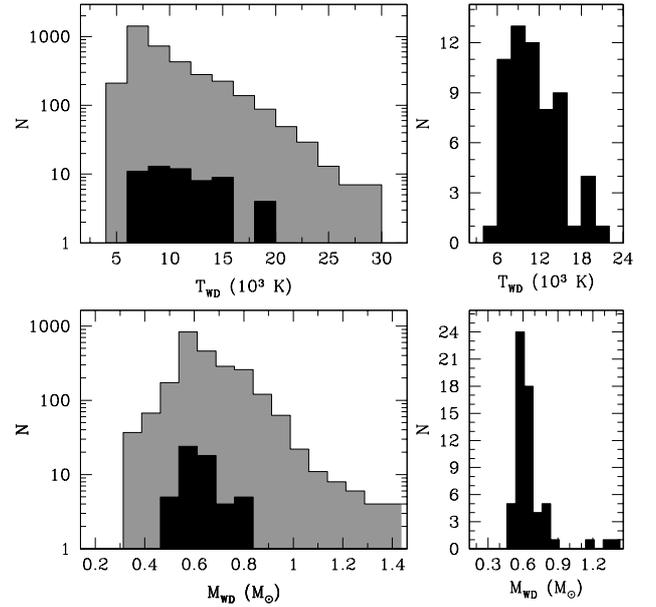}
    \caption{Left   panels:   WD   effective  temperature   and   mass
      distributions in logarithmic  scale. In gray we  show the entire
      100\,pc sample of {\it Gaia} WDs with reliable IR photometry, in
      black  those  WD   candidates  to  show  IR  excess   due  to  a
      circumstellar                                               disk
      (Tables\,\ref{t:dusty}--\ref{t:dusty2}).   Right   panels:   the
      parameter distributions in linear scale  of the WD candidates to
      show IR excess due to the  presence of a cirsumstellar disk. The
      WD masses above 1\Msun should  be taken with caution, especially
      for those objects displaying IR excess, since these are possibly
      non-DA WDs.}
    \label{fig:histo_wd}
\end{figure}

The effective  temperature and  mass distributions for  the 60  WDs in
which the IR  excess is confirmed/expected to arise  from the presence
of a circumstellar  dust disk (indicated in bold face  in the last two
columns  of Tables\,\ref{t:dusty}-\ref{t:dusty2})  are illustrated  in
Fig.\,\ref{fig:histo_wd}.   Inspection  of   the  Figure  reveals  the
effective temperatures are  concentrated between 6,000-20,000\,K, with
a peak at $\simeq$9,000\,K. No WDs are found below 6,000\,K due to the
$G_{\rm   BP}-G_{\rm   RP}   <$   0.8    mag   cut   we   imposed   in
Section\,\ref{Iden}.   The  drop  of  IR-excess  WD  candidates  above
$\simeq$20,000\,K    is     in    line    with     previous    studies
\citep[e.g.][]{Bergfors2014,   Barber14,   Barber16}.   The   physical
mechanism causing  this could  be that the  accretion of  the material
onto the  WD is supplied  by a pure gas  disc, which results  from the
sublimation  of  optically  thin  dust   due  to  the  high  effective
temperatures \citep{Bonsor2017}.  A drop  in the fraction of IR-excess
WDs below  $\simeq$8,000--10,000,K is also reported  in previous works
\citep[e.g.][]{Bergfors2014},  a   result  that  favours   a  positive
correlation between the WD effective  temperature and the detection of
IR-excess.  In the top-left  panel of Fig.\,\ref{fig:histo_wd} one can
clearly  see   the  same   tendency,  i.e.    the  fraction   of  cool
($\la$8,000--10,000\,K) WDs  displaying IR excess  decreases. However,
since  the {\it  Gaia} 100pc  WD  sample is  volume-limited and  hence
dominated by WDs cooler  than 10,000\,K \citep{Jimenez-Esteban18}, the
peak   of   the   effective   temperature   distribution   occurs   at
$\simeq$9,000\,K rather than at higher temperatures (see the top-right
panel of Figure\,\ref{fig:histo_wd}).

The mass  distribution of the  WD candidates for displaying  IR excess
due to a  circumstellar dust disk is clearly  dominated by $\simeq$0.6
M$_\mathrm{\odot}$ objects.  Indeed, $\simeq$50\%  of the objects have
masses between 0.55 and 0.65\,\Msun. It is worth mentioning that three
WDs have masses above 1\Msun, one  of them near the Chandrasekhar mass
limit. The  masses of  these WDs  should be  taken with  caution since
there exists the possibility these  are non-DAs. WDs more massive than
$>$0.8 M$_\mathrm{\odot}$ are not generally found to display IR excess
\citep{Mullally2007,  Barber16}.  This  is also  observed in  our mass
distribution, where only $\simeq$5\% of the objects are located in the
0.8--1.0 mass range.

Based    on     the    60    confirmed/expected    dusty     WDs    in
Tables\,\ref{t:dusty}-\ref{t:dusty2},  we  calculate   a  fraction  of
IR-excess WDs  due to the  existence of  a circumstellar dust  disk of
1.6$\pm$0.2\% \footnote{The  error is calculated  as $\sqrt{\frac{frac
      \times   \left(   1-frac  \right)   }{N_\mathrm{tot}}}$,   where
  N$_\mathrm{tot}$ is the  total number of WDs and  \emph{frac} is the
  IR-excess  fraction of  WDs.}.  As  we have  already mentioned,  the
detection of IR-excess  seems to be positively correlated  with the WD
effective temperature.   Thus, if  we exclude  all WDs  with effective
temperatures below 8,000\,K from our sample, the fraction of IR-excess
WDs  increases to  2.3$\pm$0.3\%, as  expected.  For  completeness, we
provide  the  fractions at  different  effective  temperature bins  in
Table\,\ref{t:fractions},  where  one  can see  the  values  gradually
decrease as  soon as  we move  towards cooler  effective temperatures.
The fractions  we derived are  in excellent agreement with  the 1--5\%
measured  percentages  by  other studies  \citep{Debes2011,  Barber16,
  Bonsor2017, Wilson2019}.  It has to be emphasised however that these
values should  be considered as  lower limits,  since many WDs  in our
sample with  reliable IR  photometry have no  counterparts at  near IR
wavelengths  (we  remind  the  reader that  all  our  candidates  have
reliable IR photometry  at both near and mid IR  wavelengths).  Of the
3,733  WDs  in  our  analysed  sample,  2802  have  near  and  mid  IR
counterparts. This implies a revised  overall WD IR-excess fraction of
2.1$\pm$0.3\%.  Moreover, if we take  into account that we assumed our
IR-excess  sample to  be 83\%  complete (Section\,\ref{method}),  this
implies a completeness-corrected fraction of 2.6$\pm$0.3\%.

\begin{table}
\centering
\caption{The  fraction of  IR-excess WDs  due  to the  existence of  a
  circumstellar dust disk for different effective temperature bins.}
\label{t:fractions}
\setlength{\tabcolsep}{1.6ex}
\begin{tabular}{ccc} 
\hline
T$_\mathrm{eff}$ range (K) & fraction (\%) & error (\%) \\
\hline
 6,000-8,000   & 0.8 & 0.2 \\
 8,000-10,000  & 1.8 & 0.5 \\
10,000-12,000  & 2.8 & 0.8 \\
12,000-14,000  & 2.9 & 1.0 \\
14,000-16,000  & 4.0 & 1.3 \\
16,000-18,000  & 0.7 & 0.7 \\
18,000-20,000  & 4.5 & 2.2 \\
20,000-22,000  & 2.0 & 2.0 \\
\hline
\end{tabular}
\end{table}

Our sample of 77 IR-excess  WD candidates contains two confirmed WD+BD
binaries (LSPM\,J0135+1445  and GD\,1400) and  two more that  are also
likely  to  harbour  brown  dwarf  companions:  PM\,J04339+2827  (Gaia
ID$=$54901403567006805)   and  549014035670068057   (yellow  dots   in
Fig.\,\ref{fig:excess}).  We  thus derive a fraction  of IR-excess WDs
due to  brown dwarf companions  of 0.10$\pm$0.05\%. If we  assume that
the  IR  excess   of  all  WDs  expected  to   harbour  companions  in
Tables\,\ref{t:dusty}-\ref{t:dusty2} is  due to  a brown  dwarf rather
than   due  to   a  low-mass   star,  the   percentage  increases   to
0.20$\pm$0.05\%,  or  0.23$\pm$0.05\%  if  we take  into  account  the
incompleteness of our  sample.  Our results are in  agreement with the
expected value of $<$0.5\% claimed by \citet{Farihi2005}.

Finally, it is worth mentioning that all IR-excess WD candidates found
in this study are classified to  be members of the Galactic thin disk,
according to  the Random Forest identification  algorithm we presented
in \citet{Torres2019}.   The proportion of  WDs within the  {\it Gaia}
100\,pc  sample belonging  to the  thin/thick disk  within the  colour
limits considered ($G_{\rm BP}-G_{\rm RP}$<0.8  mag) is expected to be
96.5:3.5 \citep{Torres2019}, which translates  into an expected number
of $\simeq$74 thin  disk and $\simeq$3 thick disk  IR-excess WDs among
our 77 identified objects. Thus, the expected number of thick disk WDs
is at  odds with  the observed  value. Given that  thick disk  WDs are
generally old ($\ga$9\,Gyr), and taking into account that the fraction
of IR-excess WDs drops for cooler (hence generally older) objects, the
discrepancy between the expected and the observed number of thin/thick
disk IR-excess WDs seems to be naturally explained.

\section{Conclusions}

We have  analysed the  SEDs of  3,733 WDs within  100pc from  the {\it
  Gaia}  volume-limited   sample  of   \citet{Jimenez-Esteban18}  with
reliable IR photometry and with  $G_{\rm BP}-G_{\rm RP}$ colours below
0.8 mag with the aim of detecting IR excess candidates. The search has
resulted in 77 identifications, 52 of which are new. 33 additional WDs
have been also identified  as potential IR-excess candidates. However,
the fact that no near IR  photometry is available in these cases makes
the reliability of the dectections  less certain. We have provided the
largest  volume-limited sample  of IR  excess WD  candidates to  date,
which represents a fraction of 2.6$\pm$0.3\% of the sample analysed in
this work. Having this large number  of WD candidates for harbouring a
circumstellar disk  at hand opens  up the possibility  to considerably
increase  our understanding  of  the properties  of extreme  planetary
systems.  A similar exercise to the one performed in this work but for
the entire \emph{Gaia}  sample of WDs is encouraging.  Even though the
\emph{Gaia} WD catalogue  icluding WDs at all  distances is magnitude-
rather  than volume-limited,  such  an analysis  would  result in  the
identification of many additional  WDs displaying IR-excess, important
for future follow-up studies.

\section*{Acknowledgements}
This work  has made use of  data from the European  Space Agency (ESA)
mission {\it  Gaia} (\url{https://www.cosmos.esa.int/gaia}), processed
by  the {\it  Gaia}  Data Processing  and  Analysis Consortium  (DPAC,
\url{https://www.cosmos.esa.int/web/gaia/dpac/consortium}).    Funding
for the DPAC has been provided by national institutions, in particular
the  institutions   participating  in  the  {\it   Gaia}  Multilateral
Agreement. This  publication makes  use of  VOSA, developed  under the
Spanish Virtual Observatory project  supported from the Spanish MINECO
through grant AyA2017-84089. This research  has made use of the SIMBAD
database, operated at  CDS, Strasbourg, France. We  acknowledge use of
the ADS bibliographic services. This  research has made use of "Aladin
sky  atlas" developed  at CDS,  Strasbourg Observatory,  France.  This
research has  made use  of Topcat \citep{Taylor05}.   ARM acknowledges
support  from  the   MINECO  under  the  Ram\'on   y  Cajal  programme
(RYC-2016-20254).    ARM  and   ST   acknowledge   support  from   the
AYA2017-86274-P  grant  and  the  AGAUR  grant  SGR-661/2017.   F.J.E.
acknowledges   financial   support   from  the   Spacetec-CM   project
(S2013/ICE-2822),    and    from    ASTERICS    project    (ID:653477,
H2020-EU.1.4.1.1.     -    Developing   new    world-class    research
infrastructures).  We   thank  the   anonymous  referee   for  her/his
suggestions and comments.

\appendix
\section{Online catalogue service}
\label{Append}

In order to help the astronomical  community on using the catalogue of
WDs with infrared excess identified in this work, we have developed an
archive     system     that     can     be     accessed     from     a
webpage\footnote{http://svo2.cab.inta-csic.es/vocats/v2/wdw3}       or
through               a              Virtual               Observatory
ConeSearch\footnote{e.g. http://svo2.cab.inta-csic.es/vocats/v2/wdw3/cs.php?RA=1.895\&DEC=-16.092\&SR=0.1\&VERB=2}.

The  archive system  implements a  very simple  search interface  that
permits queries  by coordinates  and range of  effective temperatures,
surface gravities  and masses.  The user can  also select  the maximum
number of sources to return (with values from 10 to unlimited).

The result of the query is a  HTML table with all the sources found in
the archive  fulfilling the  search criteria. The  result can  also be
downloaded as  a VOTable or  a CSV  file. Detailed information  on the
output fields can be obtained placing the mouse over the question mark
(``?")  located close  to the  name of  the column.  The archive  also
implements    the   SAMP\footnote{http://www.ivoa.net/documents/SAMP/}
(Simple  Application  Messaging)  Virtual Observatory  protocol.  SAMP
allows Virtual Observatory applications to communicate with each other
in  a seamless  and transparent  manner for  the user.  This way,  the
results of a query can be easily transferred to other VO applications,
such as, for instance, Topcat.




\begin{thebibliography}{}
\makeatletter
\relax
\def\mn@urlcharsother{\let\do\@makeother \do\$\do\&\do\#\do\^\do\_\do\%\do\~}
\def\mn@doi{\begingroup\mn@urlcharsother \@ifnextchar [ {\mn@doi@}
  {\mn@doi@[]}}
\def\mn@doi@[#1]#2{\def\@tempa{#1}\ifx\@tempa\@empty \href
  {http://dx.doi.org/#2} {doi:#2}\else \href {http://dx.doi.org/#2} {#1}\fi
  \endgroup}
\def\mn@eprint#1#2{\mn@eprint@#1:#2::\@nil}
\def\mn@eprint@arXiv#1{\href {http://arxiv.org/abs/#1} {{\tt arXiv:#1}}}
\def\mn@eprint@dblp#1{\href {http://dblp.uni-trier.de/rec/bibtex/#1.xml}
  {dblp:#1}}
\def\mn@eprint@#1:#2:#3:#4\@nil{\def\@tempa {#1}\def\@tempb {#2}\def\@tempc
  {#3}\ifx \@tempc \@empty \let \@tempc \@tempb \let \@tempb \@tempa \fi \ifx
  \@tempb \@empty \def\@tempb {arXiv}\fi \@ifundefined
  {mn@eprint@\@tempb}{\@tempb:\@tempc}{\expandafter \expandafter \csname
  mn@eprint@\@tempb\endcsname \expandafter{\@tempc}}}

\bibitem[\protect\citeauthoryear{{Alam} et~al.,}{{Alam} et~al.}{2015}]{Alam15}
{Alam} S.,  et~al., 2015, \mn@doi [\apjs] {10.1088/0067-0049/219/1/12}, \href
  {http://cdsads.u-strasbg.fr/abs/2015ApJS..219...12A} {219, 12}

\bibitem[\protect\citeauthoryear{{Allard}, {Homeier}  \& {Freytag}}{{Allard}
  et~al.}{2012}]{Allard12}
{Allard} F.,  {Homeier} D.,   {Freytag} B.,  2012, \mn@doi [Philosophical
  Transactions of the Royal Society of London Series A]
  {10.1098/rsta.2011.0269}, \href
  {https://ui.adsabs.harvard.edu/#abs/2012RSPTA.370.2765A} {370, 2765}

\bibitem[\protect\citeauthoryear{{Althaus}, {C{\'o}rsico}, {Isern}  \&
  {Garc{\'{\i}}a-Berro}}{{Althaus} et~al.}{2010}]{Althaus2010a}
{Althaus} L.~G.,  {C{\'o}rsico} A.~H.,  {Isern} J.,   {Garc{\'{\i}}a-Berro} E.,
   2010, \mn@doi [\aapr] {10.1007/s00159-010-0033-1}, \href
  {http://cdsads.u-strasbg.fr/abs/2010A\%26ARv..18..471A} {18, 471}

\bibitem[\protect\citeauthoryear{{Badenes}, {van Kerkwijk}, {Kilic},
  {Bickerton}, {Mazeh}, {Mullally}, {Tal-Or}  \& {Thompson}}{{Badenes}
  et~al.}{2013}]{Badenes2013}
{Badenes} C.,  {van Kerkwijk} M.~H.,  {Kilic} M.,  {Bickerton} S.~J.,  {Mazeh}
  T.,  {Mullally} F.,  {Tal-Or} L.,   {Thompson} S.~E.,  2013, \mn@doi [\mnras]
  {10.1093/mnras/sts646}, \href
  {http://adsabs.harvard.edu/abs/2013MNRAS.429.3596B} {429, 3596}

\bibitem[\protect\citeauthoryear{{Barber}, {Patterson}, {Kilic}, {Leggett},
  {Dufour}, {Bloom}  \& {Starr}}{{Barber} et~al.}{2012}]{Barber2012}
{Barber} S.~D.,  {Patterson} A.~J.,  {Kilic} M.,  {Leggett} S.~K.,  {Dufour}
  P.,  {Bloom} J.~S.,   {Starr} D.~L.,  2012, \mn@doi [\apj]
  {10.1088/0004-637X/760/1/26}, \href
  {http://adsabs.harvard.edu/abs/2012ApJ...760...26B} {760, 26}

\bibitem[\protect\citeauthoryear{{Barber}, {Kilic}, {Brown}  \&
  {Gianninas}}{{Barber} et~al.}{2014}]{Barber14}
{Barber} S.~D.,  {Kilic} M.,  {Brown} W.~R.,   {Gianninas} A.,  2014, \mn@doi
  [\apj] {10.1088/0004-637X/786/2/77}, \href
  {https://ui.adsabs.harvard.edu/\#abs/2014ApJ...786...77B} {786, 77}

\bibitem[\protect\citeauthoryear{{Barber}, {Belardi}, {Kilic}  \&
  {Gianninas}}{{Barber} et~al.}{2016}]{Barber16}
{Barber} S.~D.,  {Belardi} C.,  {Kilic} M.,   {Gianninas} A.,  2016, \mn@doi
  [\mnras] {10.1093/mnras/stw683}, \href
  {https://ui.adsabs.harvard.edu/#abs/2016MNRAS.459.1415B} {459, 1415}

\bibitem[\protect\citeauthoryear{{Bayo}, {Rodrigo}, {Barrado y Navascu{\'e}s},
  {Solano}, {Guti{\'e}rrez}, {Morales-Calder{\'o}n}  \& {Allard}}{{Bayo}
  et~al.}{2008}]{Bayo08}
{Bayo} A.,  {Rodrigo} C.,  {Barrado y Navascu{\'e}s} D.,  {Solano} E.,
  {Guti{\'e}rrez} R.,  {Morales-Calder{\'o}n} M.,   {Allard} F.,  2008, \mn@doi
  [\aap] {10.1051/0004-6361:200810395}, \href
  {http://adsabs.harvard.edu/abs/2008A\%26A...492..277B} {492, 277}

\bibitem[\protect\citeauthoryear{{Bergeron} et~al.,}{{Bergeron}
  et~al.}{2011}]{Bergeron2011}
{Bergeron} P.,  et~al., 2011, \mn@doi [\apj] {10.1088/0004-637X/737/1/28},
  \href {http://adsabs.harvard.edu/abs/2011ApJ...737...28B} {737, 28}

\bibitem[\protect\citeauthoryear{{Bergfors}, {Farihi}, {Dufour}  \&
  {Rocchetto}}{{Bergfors} et~al.}{2014}]{Bergfors2014}
{Bergfors} C.,  {Farihi} J.,  {Dufour} P.,   {Rocchetto} M.,  2014, \mn@doi
  [\mnras] {10.1093/mnras/stu1565}, \href
  {http://adsabs.harvard.edu/abs/2014MNRAS.444.2147B} {444, 2147}

\bibitem[\protect\citeauthoryear{{Bianchi} \& {GALEX Team}}{{Bianchi} \& {GALEX
  Team}}{2000}]{Bianchi00}
{Bianchi} L.,  {GALEX Team} 2000, \memsai, \href
  {http://adsabs.harvard.edu/abs/2000MmSAI..71.1123B} {71, 1123}

\bibitem[\protect\citeauthoryear{{Bonnarel} et~al.,}{{Bonnarel}
  et~al.}{2000}]{Bonnarel00}
{Bonnarel} F.,  et~al., 2000, \mn@doi [\aaps] {10.1051/aas:2000331}, \href
  {http://adsabs.harvard.edu/abs/2000A\%26AS..143...33B} {143, 33}

\bibitem[\protect\citeauthoryear{{Bonsor}, {Mustill}  \& {Wyatt}}{{Bonsor}
  et~al.}{2011}]{Bonsor2011}
{Bonsor} A.,  {Mustill} A.~J.,   {Wyatt} M.~C.,  2011, \mn@doi [\mnras]
  {10.1111/j.1365-2966.2011.18524.x}, \href
  {http://adsabs.harvard.edu/abs/2011MNRAS.414..930B} {414, 930}

\bibitem[\protect\citeauthoryear{{Bonsor}, {Farihi}, {Wyatt}  \& {van
  Lieshout}}{{Bonsor} et~al.}{2017}]{Bonsor2017}
{Bonsor} A.,  {Farihi} J.,  {Wyatt} M.~C.,   {van Lieshout} R.,  2017, \mn@doi
  [\mnras] {10.1093/mnras/stx425}, \href
  {http://adsabs.harvard.edu/abs/2017MNRAS.468..154B} {468, 154}

\bibitem[\protect\citeauthoryear{{Burleigh}, {Clarke}  \& {Hodgkin}}{{Burleigh}
  et~al.}{2002}]{Burleigh2002}
{Burleigh} M.~R.,  {Clarke} F.~J.,   {Hodgkin} S.~T.,  2002, \mn@doi [\mnras]
  {10.1046/j.1365-8711.2002.05417.x}, \href
  {http://adsabs.harvard.edu/abs/2002MNRAS.331L..41B} {331, L41}

\bibitem[\protect\citeauthoryear{{Castanheira} \& {Kepler}}{{Castanheira} \&
  {Kepler}}{2008}]{Castanheira2008}
{Castanheira} B.~G.,  {Kepler} S.~O.,  2008, \mn@doi [\mnras]
  {10.1111/j.1365-2966.2008.12851.x}, \href
  {http://adsabs.harvard.edu/abs/2008MNRAS.385..430C} {385, 430}

\bibitem[\protect\citeauthoryear{{Chambers} et~al.,}{{Chambers}
  et~al.}{2016}]{Chambers16}
{Chambers} K.~C.,  et~al., 2016, preprint, \href
  {http://cdsads.u-strasbg.fr/abs/2016arXiv161205560C} {} (\mn@eprint {arXiv}
  {1612.05560})

\bibitem[\protect\citeauthoryear{{Cross} et~al.,}{{Cross}
  et~al.}{2012}]{Cross12}
{Cross} N.~J.~G.,  et~al., 2012, \mn@doi [\aap] {10.1051/0004-6361/201219505},
  \href {http://adsabs.harvard.edu/abs/2012A\%26A...548A.119C} {548, A119}

\bibitem[\protect\citeauthoryear{{Dark Energy Survey Collaboration}
  et~al.,}{{Dark Energy Survey Collaboration} et~al.}{2016}]{DESC16}
{Dark Energy Survey Collaboration} et~al., 2016, \mn@doi [\mnras]
  {10.1093/mnras/stw641}, \href
  {http://adsabs.harvard.edu/abs/2016MNRAS.460.1270D} {460, 1270}

\bibitem[\protect\citeauthoryear{{Debes} \& {Sigurdsson}}{{Debes} \&
  {Sigurdsson}}{2002}]{Debes2002}
{Debes} J.~H.,  {Sigurdsson} S.,  2002, \mn@doi [\apj] {10.1086/340291}, \href
  {http://adsabs.harvard.edu/abs/2002ApJ...572..556D} {572, 556}

\bibitem[\protect\citeauthoryear{{Debes}, {Hoard}, {Wachter}, {Leisawitz}  \&
  {Cohen}}{{Debes} et~al.}{2011}]{Debes2011}
{Debes} J.~H.,  {Hoard} D.~W.,  {Wachter} S.,  {Leisawitz} D.~T.,   {Cohen} M.,
   2011, \mn@doi [\apjs] {10.1088/0067-0049/197/2/38}, \href
  {http://adsabs.harvard.edu/abs/2011ApJS..197...38D} {197, 38}

\bibitem[\protect\citeauthoryear{{Debes}, {Walsh}  \& {Stark}}{{Debes}
  et~al.}{2012}]{Debes2012}
{Debes} J.~H.,  {Walsh} K.~J.,   {Stark} C.,  2012, \mn@doi [\apj]
  {10.1088/0004-637X/747/2/148}, \href
  {http://adsabs.harvard.edu/abs/2012ApJ...747..148D} {747, 148}

\bibitem[\protect\citeauthoryear{{Dennihy}, {Debes}, {Dunlap}, {Dufour},
  {Teske}  \& {Clemens}}{{Dennihy} et~al.}{2016}]{Dennihy16}
{Dennihy} E.,  {Debes} J.~H.,  {Dunlap} B.~H.,  {Dufour} P.,  {Teske} J.~K.,
  {Clemens} J.~C.,  2016, \mn@doi [\apj] {10.3847/0004-637X/831/1/31}, \href
  {https://ui.adsabs.harvard.edu/#abs/2016ApJ...831...31D} {831, 31}

\bibitem[\protect\citeauthoryear{{Dennihy}, {Clemens}, {Debes}, {Dunlap},
  {Kilkenny}, {O'Brien}  \& {Fuchs}}{{Dennihy} et~al.}{2017}]{Dennihy17}
{Dennihy} E.,  {Clemens} J.~C.,  {Debes} J.~H.,  {Dunlap} B.~H.,  {Kilkenny}
  D.,  {O'Brien} P.~C.,   {Fuchs} J.~T.,  2017, \mn@doi [\apj]
  {10.3847/1538-4357/aa8ef2}, \href
  {https://ui.adsabs.harvard.edu/\#abs/2017ApJ...849...77D} {849, 77}

\bibitem[\protect\citeauthoryear{{Dennihy}, {Clemens}, {Dunlap}, {Fanale},
  {Fuchs}  \& {Hermes}}{{Dennihy} et~al.}{2018}]{Dennihy2018}
{Dennihy} E.,  {Clemens} J.~C.,  {Dunlap} B.~H.,  {Fanale} S.~M.,  {Fuchs}
  J.~T.,   {Hermes} J.~J.,  2018, \mn@doi [\apj] {10.3847/1538-4357/aaa89b},
  \href {https://ui.adsabs.harvard.edu/abs/2018ApJ...854...40D} {854, 40}

\bibitem[\protect\citeauthoryear{{El-Badry} \& {Rix}}{{El-Badry} \&
  {Rix}}{2018}]{Badry2018}
{El-Badry} K.,  {Rix} H.-W.,  2018, \mn@doi [\mnras] {10.1093/mnras/sty2186},
  \href {http://adsabs.harvard.edu/abs/2018MNRAS.480.4884E} {480, 4884}

\bibitem[\protect\citeauthoryear{{Farihi}}{{Farihi}}{2016}]{Farihi2016}
{Farihi} J.,  2016, \mn@doi [\nar] {10.1016/j.newar.2016.03.001}, \href
  {https://ui.adsabs.harvard.edu/abs/2016NewAR..71....9F} {71, 9}

\bibitem[\protect\citeauthoryear{{Farihi} \& {Christopher}}{{Farihi} \&
  {Christopher}}{2004}]{Farihi2004}
{Farihi} J.,  {Christopher} M.,  2004, \mn@doi [\aj] {10.1086/423919}, \href
  {http://adsabs.harvard.edu/abs/2004AJ....128.1868F} {128, 1868}

\bibitem[\protect\citeauthoryear{{Farihi}, {Becklin}  \& {Zuckerman}}{{Farihi}
  et~al.}{2005}]{Farihi2005}
{Farihi} J.,  {Becklin} E.~E.,   {Zuckerman} B.,  2005, \mn@doi [\apjs]
  {10.1086/444362}, \href {http://adsabs.harvard.edu/abs/2005ApJS..161..394F}
  {161, 394}

\bibitem[\protect\citeauthoryear{{Farihi}, {Zuckerman}  \& {Becklin}}{{Farihi}
  et~al.}{2008a}]{Farihi08b}
{Farihi} J.,  {Zuckerman} B.,   {Becklin} E.~E.,  2008a, \mn@doi [\apj]
  {10.1086/521715}, \href
  {https://ui.adsabs.harvard.edu/\#abs/2008ApJ...674..431F} {674, 431}

\bibitem[\protect\citeauthoryear{{Farihi}, {Becklin}  \& {Zuckerman}}{{Farihi}
  et~al.}{2008b}]{Farihi08}
{Farihi} J.,  {Becklin} E.~E.,   {Zuckerman} B.,  2008b, \mn@doi [\apj]
  {10.1086/588726}, \href
  {https://ui.adsabs.harvard.edu/\#abs/2008ApJ...681.1470F} {681, 1470}

\bibitem[\protect\citeauthoryear{{Farihi}, {Jura}  \& {Zuckerman}}{{Farihi}
  et~al.}{2009}]{Farihi09}
{Farihi} J.,  {Jura} M.,   {Zuckerman} B.,  2009, \mn@doi [\apj]
  {10.1088/0004-637X/694/2/805}, \href
  {https://ui.adsabs.harvard.edu/\#abs/2009ApJ...694..805F} {694, 805}

\bibitem[\protect\citeauthoryear{{Farihi}, {Jura}, {Lee}  \&
  {Zuckerman}}{{Farihi} et~al.}{2010}]{Farihi10}
{Farihi} J.,  {Jura} M.,  {Lee} J.~E.,   {Zuckerman} B.,  2010, \mn@doi [\apj]
  {10.1088/0004-637X/714/2/1386}, \href
  {https://ui.adsabs.harvard.edu/\#abs/2010ApJ...714.1386F} {714, 1386}

\bibitem[\protect\citeauthoryear{{Farihi}, {Burleigh}, {Holberg}, {Casewell}
  \& {Barstow}}{{Farihi} et~al.}{2011}]{Farihi2011}
{Farihi} J.,  {Burleigh} M.~R.,  {Holberg} J.~B.,  {Casewell} S.~L.,
  {Barstow} M.~A.,  2011, \mn@doi [\mnras] {10.1111/j.1365-2966.2011.19354.x},
  \href {http://adsabs.harvard.edu/abs/2011MNRAS.417.1735F} {417, 1735}

\bibitem[\protect\citeauthoryear{{Farihi}, {G{\"a}nsicke}  \&
  {Koester}}{{Farihi} et~al.}{2013}]{Farihi2013}
{Farihi} J.,  {G{\"a}nsicke} B.~T.,   {Koester} D.,  2013, \mn@doi [Science]
  {10.1126/science.1239447}, \href
  {https://ui.adsabs.harvard.edu/abs/2013Sci...342..218F} {342, 218}

\bibitem[\protect\citeauthoryear{{Gaia Collaboration}, {Brown}, {Vallenari},
  {Prusti}, {de Bruijne}, {Babusiaux}  \& {Bailer-Jones}}{{Gaia Collaboration}
  et~al.}{2018}]{Brown18}
{Gaia Collaboration} {Brown} A.~G.~A.,  {Vallenari} A.,  {Prusti} T.,  {de
  Bruijne} J.~H.~J.,  {Babusiaux} C.,   {Bailer-Jones} C.~A.~L.,  2018,
  preprint, \href {http://cdsads.u-strasbg.fr/abs/2018arXiv180409365G} {}
  (\mn@eprint {arXiv} {1804.09365})

\bibitem[\protect\citeauthoryear{{G{\"a}nsicke}, {Marsh}, {Southworth}  \&
  {Rebassa-Mansergas}}{{G{\"a}nsicke} et~al.}{2006}]{Gaensicke2006}
{G{\"a}nsicke} B.~T.,  {Marsh} T.~R.,  {Southworth} J.,   {Rebassa-Mansergas}
  A.,  2006, \mn@doi [Science] {10.1126/science.1135033}, \href
  {http://adsabs.harvard.edu/abs/2006Sci...314.1908G} {314, 1908}

\bibitem[\protect\citeauthoryear{{G{\"a}nsicke}, {Koester}, {Farihi}, {Girven},
  {Parsons}  \& {Breedt}}{{G{\"a}nsicke} et~al.}{2012}]{Gaensicke2012}
{G{\"a}nsicke} B.~T.,  {Koester} D.,  {Farihi} J.,  {Girven} J.,  {Parsons}
  S.~G.,   {Breedt} E.,  2012, \mn@doi [\mnras]
  {10.1111/j.1365-2966.2012.21201.x}, \href
  {http://adsabs.harvard.edu/abs/2012MNRAS.424..333G} {424, 333}

\bibitem[\protect\citeauthoryear{{Gentile Fusillo} et~al.,}{{Gentile Fusillo}
  et~al.}{2019}]{Gentile2019}
{Gentile Fusillo} N.~P.,  et~al., 2019, \mn@doi [\mnras]
  {10.1093/mnras/sty3016}, \href
  {http://adsabs.harvard.edu/abs/2019MNRAS.482.4570G} {482, 4570}

\bibitem[\protect\citeauthoryear{{Gianninas}, {Bergeron}  \&
  {Ruiz}}{{Gianninas} et~al.}{2011}]{Gianninas2011}
{Gianninas} A.,  {Bergeron} P.,   {Ruiz} M.~T.,  2011, \mn@doi [\apj]
  {10.1088/0004-637X/743/2/138}, \href
  {http://adsabs.harvard.edu/abs/2011ApJ...743..138G} {743, 138}

\bibitem[\protect\citeauthoryear{{Girven}, {Brinkworth}, {Farihi},
  {G{\"a}nsicke}, {Hoard}, {Marsh}  \& {Koester}}{{Girven}
  et~al.}{2012}]{Girven12}
{Girven} J.,  {Brinkworth} C.~S.,  {Farihi} J.,  {G{\"a}nsicke} B.~T.,  {Hoard}
  D.~W.,  {Marsh} T.~R.,   {Koester} D.,  2012, \mn@doi [\apj]
  {10.1088/0004-637X/749/2/154}, \href
  {https://ui.adsabs.harvard.edu/\#abs/2012ApJ...749..154G} {749, 154}

\bibitem[\protect\citeauthoryear{{Hewett}, {Warren}, {Leggett}  \&
  {Hodgkin}}{{Hewett} et~al.}{2006}]{Hewett06}
{Hewett} P.~C.,  {Warren} S.~J.,  {Leggett} S.~K.,   {Hodgkin} S.~T.,  2006,
  \mn@doi [\mnras] {10.1111/j.1365-2966.2005.09969.x}, \href
  {http://adsabs.harvard.edu/abs/2006MNRAS.367..454H} {367, 454}

\bibitem[\protect\citeauthoryear{{Hoard}, {Debes}, {Wachter}, {Leisawitz}  \&
  {Cohen}}{{Hoard} et~al.}{2013}]{Hoard13}
{Hoard} D.~W.,  {Debes} J.~H.,  {Wachter} S.,  {Leisawitz} D.~T.,   {Cohen} M.,
   2013, \mn@doi [\apj] {10.1088/0004-637X/770/1/21}, \href
  {https://ui.adsabs.harvard.edu/#abs/2013ApJ...770...21H} {770, 21}

\bibitem[\protect\citeauthoryear{{Hollands}, {G{\"a}nsicke}  \&
  {Koester}}{{Hollands} et~al.}{2018a}]{Hollands2018b}
{Hollands} M.~A.,  {G{\"a}nsicke} B.~T.,   {Koester} D.,  2018a, \mn@doi
  [\mnras] {10.1093/mnras/sty592}, \href
  {https://ui.adsabs.harvard.edu/abs/2018MNRAS.477...93H} {477, 93}

\bibitem[\protect\citeauthoryear{{Hollands}, {Tremblay}, {G{\"a}nsicke},
  {Gentile-Fusillo}  \& {Toonen}}{{Hollands} et~al.}{2018b}]{Hollands2018}
{Hollands} M.~A.,  {Tremblay} P.-E.,  {G{\"a}nsicke} B.~T.,  {Gentile-Fusillo}
  N.~P.,   {Toonen} S.,  2018b, \mn@doi [\mnras] {10.1093/mnras/sty2057}, \href
  {http://adsabs.harvard.edu/abs/2018MNRAS.480.3942H} {480, 3942}

\bibitem[\protect\citeauthoryear{{Jim{\'e}nez-Esteban}, {Torres},
  {Rebassa-Mansergas}, {Skorobogatov}, {Solano}, {Cantero}  \&
  {Rodrigo}}{{Jim{\'e}nez-Esteban} et~al.}{2018}]{Jimenez-Esteban18}
{Jim{\'e}nez-Esteban} F.~M.,  {Torres} S.,  {Rebassa-Mansergas} A.,
  {Skorobogatov} G.,  {Solano} E.,  {Cantero} C.,   {Rodrigo} C.,  2018,
  \mn@doi [\mnras] {10.1093/mnras/sty2120}, \href
  {http://cdsads.u-strasbg.fr/abs/2018MNRAS.480.4505J} {480, 4505}

\bibitem[\protect\citeauthoryear{{Jura}}{{Jura}}{2008}]{Jura2008}
{Jura} M.,  2008, \mn@doi [\aj] {10.1088/0004-6256/135/5/1785}, \href
  {http://adsabs.harvard.edu/abs/2008AJ....135.1785J} {135, 1785}

\bibitem[\protect\citeauthoryear{{Jura}, {Farihi}  \& {Zuckerman}}{{Jura}
  et~al.}{2007}]{Jura07}
{Jura} M.,  {Farihi} J.,   {Zuckerman} B.,  2007, \mn@doi [\apj]
  {10.1086/518767}, \href
  {https://ui.adsabs.harvard.edu/\#abs/2007ApJ...663.1285J} {663, 1285}

\bibitem[\protect\citeauthoryear{{Kilic}, {Patterson}, {Barber}, {Leggett}  \&
  {Dufour}}{{Kilic} et~al.}{2012}]{Kilic12}
{Kilic} M.,  {Patterson} A.~J.,  {Barber} S.,  {Leggett} S.~K.,   {Dufour} P.,
  2012, \mn@doi [\mnras] {10.1111/j.1745-3933.2011.01177.x}, \href
  {https://ui.adsabs.harvard.edu/\#abs/2012MNRAS.419L..59K} {419, L59}

\bibitem[\protect\citeauthoryear{{Klein}, {Jura}, {Koester}, {Zuckerman}  \&
  {Melis}}{{Klein} et~al.}{2010}]{Klein2010}
{Klein} B.,  {Jura} M.,  {Koester} D.,  {Zuckerman} B.,   {Melis} C.,  2010,
  \mn@doi [\apj] {10.1088/0004-637X/709/2/950}, \href
  {http://adsabs.harvard.edu/abs/2010ApJ...709..950K} {709, 950}

\bibitem[\protect\citeauthoryear{{Koester}}{{Koester}}{2010}]{Koester10}
{Koester} D.,  2010, \memsai, \href
  {http://adsabs.harvard.edu/abs/2010MmSAI..81..921K} {81, 921}

\bibitem[\protect\citeauthoryear{{Koester} \& {Kepler}}{{Koester} \&
  {Kepler}}{2015}]{Koester2015}
{Koester} D.,  {Kepler} S.~O.,  2015, \mn@doi [\aap]
  {10.1051/0004-6361/201527169}, \href
  {http://adsabs.harvard.edu/abs/2015A\%26A...583A..86K} {583, A86}

\bibitem[\protect\citeauthoryear{{Koester}, {G{\"a}nsicke}  \&
  {Farihi}}{{Koester} et~al.}{2014}]{Koester2014}
{Koester} D.,  {G{\"a}nsicke} B.~T.,   {Farihi} J.,  2014, \mn@doi [\aap]
  {10.1051/0004-6361/201423691}, \href
  {http://adsabs.harvard.edu/abs/2014A\%26A...566A..34K} {566, A34}

\bibitem[\protect\citeauthoryear{{Lada} et~al.,}{{Lada} et~al.}{2006}]{Lada06}
{Lada} C.~J.,  et~al., 2006, \mn@doi [\aj] {10.1086/499808}, \href
  {http://adsabs.harvard.edu/abs/2006AJ....131.1574L} {131, 1574}

\bibitem[\protect\citeauthoryear{{Limoges}, {Bergeron}  \&
  {L{\'e}pine}}{{Limoges} et~al.}{2015}]{Limoges2015}
{Limoges} M.-M.,  {Bergeron} P.,   {L{\'e}pine} S.,  2015, \mn@doi [\apjs]
  {10.1088/0067-0049/219/2/19}, \href
  {http://adsabs.harvard.edu/abs/2015ApJS..219...19L} {219, 19}

\bibitem[\protect\citeauthoryear{{Mullally}, {Kilic}, {Reach}, {Kuchner}, {von
  Hippel}, {Burrows}  \& {Winget}}{{Mullally} et~al.}{2007}]{Mullally2007}
{Mullally} F.,  {Kilic} M.,  {Reach} W.~T.,  {Kuchner} M.~J.,  {von Hippel} T.,
   {Burrows} A.,   {Winget} D.~E.,  2007, \mn@doi [\apjs] {10.1086/511858},
  \href {http://adsabs.harvard.edu/abs/2007ApJS..171..206M} {171, 206}

\bibitem[\protect\citeauthoryear{{Pecaut} \& {Mamajek}}{{Pecaut} \&
  {Mamajek}}{2013}]{Pecaut2013}
{Pecaut} M.~J.,  {Mamajek} E.~E.,  2013, \mn@doi [\apjs]
  {10.1088/0067-0049/208/1/9}, \href
  {http://adsabs.harvard.edu/abs/2013ApJS..208....9P} {208, 9}

\bibitem[\protect\citeauthoryear{{Raddi}, {G{\"a}nsicke}, {Koester}, {Farihi},
  {Hermes}, {Scaringi}, {Breedt}  \& {Girven}}{{Raddi}
  et~al.}{2015}]{Raddi2015}
{Raddi} R.,  {G{\"a}nsicke} B.~T.,  {Koester} D.,  {Farihi} J.,  {Hermes}
  J.~J.,  {Scaringi} S.,  {Breedt} E.,   {Girven} J.,  2015, \mn@doi [\mnras]
  {10.1093/mnras/stv701}, \href
  {https://ui.adsabs.harvard.edu/abs/2015MNRAS.450.2083R} {450, 2083}

\bibitem[\protect\citeauthoryear{{Reach}, {Kuchner}, {von Hippel}, {Burrows},
  {Mullally}, {Kilic}  \& {Winget}}{{Reach} et~al.}{2005}]{Reach05}
{Reach} W.~T.,  {Kuchner} M.~J.,  {von Hippel} T.,  {Burrows} A.,  {Mullally}
  F.,  {Kilic} M.,   {Winget} D.~E.,  2005, \mn@doi [\apj] {10.1086/499561},
  \href {https://ui.adsabs.harvard.edu/\#abs/2005ApJ...635L.161R} {635, L161}

\bibitem[\protect\citeauthoryear{{Rebassa-Mansergas}, {G{\"a}nsicke},
  {Schreiber}, {Koester}  \& {Rodr{\'{\i}}guez-Gil}}{{Rebassa-Mansergas}
  et~al.}{2010}]{Rebassa2010}
{Rebassa-Mansergas} A.,  {G{\"a}nsicke} B.~T.,  {Schreiber} M.~R.,  {Koester}
  D.,   {Rodr{\'{\i}}guez-Gil} P.,  2010, \mn@doi [\mnras]
  {10.1111/j.1365-2966.2009.15915.x}, \href
  {http://adsabs.harvard.edu/abs/2010MNRAS.402..620R} {402, 620}

\bibitem[\protect\citeauthoryear{{Rebassa-Mansergas}, {Ren}, {Parsons},
  {G{\"a}nsicke}, {Schreiber}, {Garc{\'{\i}}a-Berro}, {Liu}  \&
  {Koester}}{{Rebassa-Mansergas} et~al.}{2016}]{Rebassa2016}
{Rebassa-Mansergas} A.,  {Ren} J.~J.,  {Parsons} S.~G.,  {G{\"a}nsicke} B.~T.,
  {Schreiber} M.~R.,  {Garc{\'{\i}}a-Berro} E.,  {Liu} X.-W.,   {Koester} D.,
  2016, \mn@doi [\mnras] {10.1093/mnras/stw554}, \href
  {http://adsabs.harvard.edu/abs/2016MNRAS.458.3808R} {458, 3808}

\bibitem[\protect\citeauthoryear{{Reiners} et~al.,}{{Reiners}
  et~al.}{2018}]{Reiners18}
{Reiners} A.,  et~al., 2018, \mn@doi [\aap] {10.1051/0004-6361/201732054},
  \href {https://ui.adsabs.harvard.edu/\#abs/2018A&A...612A..49R} {612, A49}

\bibitem[\protect\citeauthoryear{{Ren} et~al.,}{{Ren} et~al.}{2014}]{Ren2014}
{Ren} J.~J.,  et~al., 2014, \mn@doi [\aap] {10.1051/0004-6361/201423689}, \href
  {https://ui.adsabs.harvard.edu/\#abs/2014A&A...570A.107R} {570, A107}

\bibitem[\protect\citeauthoryear{{Renedo}, {Althaus}, {Miller Bertolami},
  {Romero}, {C{\'o}rsico}, {Rohrmann}  \& {Garc{\'{\i}}a-Berro}}{{Renedo}
  et~al.}{2010}]{Renedo2010}
{Renedo} I.,  {Althaus} L.~G.,  {Miller Bertolami} M.~M.,  {Romero} A.~D.,
  {C{\'o}rsico} A.~H.,  {Rohrmann} R.~D.,   {Garc{\'{\i}}a-Berro} E.,  2010,
  \mn@doi [\apj] {10.1088/0004-637X/717/1/183}, \href
  {http://adsabs.harvard.edu/abs/2010ApJ...717..183R} {717, 183}

\bibitem[\protect\citeauthoryear{{Rocchetto}, {Farihi}, {G{\"a}nsicke}  \&
  {Bergfors}}{{Rocchetto} et~al.}{2015}]{Rocchetto15}
{Rocchetto} M.,  {Farihi} J.,  {G{\"a}nsicke} B.~T.,   {Bergfors} C.,  2015,
  \mn@doi [\mnras] {10.1093/mnras/stv282}, \href
  {https://ui.adsabs.harvard.edu/#abs/2015MNRAS.449..574R} {449, 574}

\bibitem[\protect\citeauthoryear{{Siess}}{{Siess}}{2007}]{Siess2007}
{Siess} L.,  2007, \mn@doi [\aap] {10.1051/0004-6361:20078132}, \href
  {http://adsabs.harvard.edu/abs/2007A\%26A...476..893S} {476, 893}

\bibitem[\protect\citeauthoryear{{Skrutskie} et~al.,}{{Skrutskie}
  et~al.}{2006}]{Skrutskie06}
{Skrutskie} M.~F.,  et~al., 2006, \mn@doi [\aj] {10.1086/498708}, \href
  {http://adsabs.harvard.edu/abs/2006AJ....131.1163S} {131, 1163}

\bibitem[\protect\citeauthoryear{{Smart}, {Marocco}, {Caballero}, {Jones},
  {Barrado}, {Beam{\'\i}n}, {Pinfield}  \& {Sarro}}{{Smart}
  et~al.}{2017}]{Smart17}
{Smart} R.~L.,  {Marocco} F.,  {Caballero} J.~A.,  {Jones} H.~R.~A.,  {Barrado}
  D.,  {Beam{\'\i}n} J.~C.,  {Pinfield} D.~J.,   {Sarro} L.~M.,  2017, \mn@doi
  [\mnras] {10.1093/mnras/stx800}, \href
  {https://ui.adsabs.harvard.edu/\#abs/2017MNRAS.469..401S} {469, 401}

\bibitem[\protect\citeauthoryear{{Steele} et~al.,}{{Steele}
  et~al.}{2013}]{Steele13}
{Steele} P.~R.,  et~al., 2013, \mn@doi [\mnras] {10.1093/mnras/sts620}, \href
  {https://ui.adsabs.harvard.edu/\#abs/2013MNRAS.429.3492S} {429, 3492}

\bibitem[\protect\citeauthoryear{{Taylor}}{{Taylor}}{2005}]{Taylor05}
{Taylor} M.~B.,  2005, in {Shopbell} P.,  {Britton} M.,   {Ebert} R.,  eds,
  Astronomical Society of the Pacific Conference Series Vol. 347, Astronomical
  Data Analysis Software and Systems XIV. p.~29

\bibitem[\protect\citeauthoryear{{Torres}, {Cantero}, {Rebassa-Mansergas},
  {Skorobogatov}, {Jim{\'e}nez-Esteban}  \& {Solano}}{{Torres}
  et~al.}{2019}]{Torres2019}
{Torres} S.,  {Cantero} C.,  {Rebassa-Mansergas} A.,  {Skorobogatov} G.,
  {Jim{\'e}nez-Esteban} F.~M.,   {Solano} E.,  2019, arXiv e-prints, \href
  {http://adsabs.harvard.edu/abs/2019arXiv190307362T} {}

\bibitem[\protect\citeauthoryear{{Tremblay} \& {Bergeron}}{{Tremblay} \&
  {Bergeron}}{2008}]{Tremblay2008}
{Tremblay} P.-E.,  {Bergeron} P.,  2008, \mn@doi [\apj] {10.1086/524134}, \href
  {http://adsabs.harvard.edu/abs/2008ApJ...672.1144T} {672, 1144}

\bibitem[\protect\citeauthoryear{{Vanderburg} et~al.,}{{Vanderburg}
  et~al.}{2015}]{Vandergurg2015}
{Vanderburg} A.,  et~al., 2015, \mn@doi [\nat] {10.1038/nature15527}, \href
  {https://ui.adsabs.harvard.edu/abs/2015Natur.526..546V} {526, 546}

\bibitem[\protect\citeauthoryear{{Veras}, {Mustill}, {Bonsor}  \&
  {Wyatt}}{{Veras} et~al.}{2013}]{Veras2013}
{Veras} D.,  {Mustill} A.~J.,  {Bonsor} A.,   {Wyatt} M.~C.,  2013, \mn@doi
  [\mnras] {10.1093/mnras/stt289}, \href
  {http://adsabs.harvard.edu/abs/2013MNRAS.431.1686V} {431, 1686}

\bibitem[\protect\citeauthoryear{{Wilson}, {G{\"a}nsicke}, {Farihi}  \&
  {Koester}}{{Wilson} et~al.}{2016}]{wilson2016}
{Wilson} D.~J.,  {G{\"a}nsicke} B.~T.,  {Farihi} J.,   {Koester} D.,  2016,
  \mn@doi [\mnras] {10.1093/mnras/stw844}, \href
  {https://ui.adsabs.harvard.edu/abs/2016MNRAS.459.3282W} {459, 3282}

\bibitem[\protect\citeauthoryear{{Wilson}, {Farihi}, {G{\"a}nsicke}  \&
  {Swan}}{{Wilson} et~al.}{2019}]{Wilson2019}
{Wilson} T.~G.,  {Farihi} J.,  {G{\"a}nsicke} B.~T.,   {Swan} A.,  2019,
  \mn@doi [\mnras] {10.1093/mnras/stz1050}, \href
  {http://adsabs.harvard.edu/abs/2019MNRAS.tmp.1000W} {}

\bibitem[\protect\citeauthoryear{{Wright} et~al.,}{{Wright}
  et~al.}{2010}]{Wright10}
{Wright} E.~L.,  et~al., 2010, \mn@doi [\aj] {10.1088/0004-6256/140/6/1868},
  \href {http://adsabs.harvard.edu/abs/2010AJ....140.1868W} {140, 1868}

\bibitem[\protect\citeauthoryear{{Wu}, {Roby}  \& {Ly}}{{Wu}
  et~al.}{2010}]{Wu10}
{Wu} X.,  {Roby} T.,   {Ly} L.,  2010, in {Mizumoto} Y.,  {Morita} K.~I.,
  {Ohishi} M.,  eds,  Astronomical Society of the Pacific Conference Series
  Vol. 434, Astronomical Data Analysis Software and Systems XIX. p.~14

\bibitem[\protect\citeauthoryear{{Xu} \& {Jura}}{{Xu} \& {Jura}}{2012}]{Xu2012}
{Xu} S.,  {Jura} M.,  2012, \mn@doi [\apj] {10.1088/0004-637X/745/1/88}, \href
  {http://adsabs.harvard.edu/abs/2012ApJ...745...88X} {745, 88}

\bibitem[\protect\citeauthoryear{{Xu} \& {Jura}}{{Xu} \&
  {Jura}}{2014}]{xu+jura2014}
{Xu} S.,  {Jura} M.,  2014, \mn@doi [\apjl] {10.1088/2041-8205/792/2/L39},
  \href {http://adsabs.harvard.edu/abs/2014ApJ...792L..39X} {792, L39}

\bibitem[\protect\citeauthoryear{{Xu}, {Jura}, {Koester}, {Klein}  \&
  {Zuckerman}}{{Xu} et~al.}{2014}]{Xu2014}
{Xu} S.,  {Jura} M.,  {Koester} D.,  {Klein} B.,   {Zuckerman} B.,  2014,
  \mn@doi [\apj] {10.1088/0004-637X/783/2/79}, \href
  {http://adsabs.harvard.edu/abs/2014ApJ...783...79X} {783, 79}

\bibitem[\protect\citeauthoryear{{Xu}, {Jura}, {Pantoja}, {Klein}, {Zuckerman},
  {Su}  \& {Meng}}{{Xu} et~al.}{2015}]{Xu2015}
{Xu} S.,  {Jura} M.,  {Pantoja} B.,  {Klein} B.,  {Zuckerman} B.,  {Su}
  K.~Y.~L.,   {Meng} H.~Y.~A.,  2015, \mn@doi [\apj]
  {10.1088/2041-8205/806/1/L5}, \href
  {https://ui.adsabs.harvard.edu/\#abs/2015ApJ...806L...5X} {806, L5}

\bibitem[\protect\citeauthoryear{{Xu}, {Zuckerman}, {Dufour}, {Young}, {Klein}
  \& {Jura}}{{Xu} et~al.}{2017}]{Xu2017}
{Xu} S.,  {Zuckerman} B.,  {Dufour} P.,  {Young} E.~D.,  {Klein} B.,   {Jura}
  M.,  2017, \mn@doi [\apjl] {10.3847/2041-8213/836/1/L7}, \href
  {https://ui.adsabs.harvard.edu/abs/2017ApJ...836L...7X} {836, L7}

\bibitem[\protect\citeauthoryear{{Zuckerman} \& {Becklin}}{{Zuckerman} \&
  {Becklin}}{1987}]{Zuckerman1987}
{Zuckerman} B.,  {Becklin} E.~E.,  1987, \mn@doi [\apjl] {10.1086/184962},
  \href {http://adsabs.harvard.edu/abs/1987ApJ...319L..99Z} {319, L99}

\bibitem[\protect\citeauthoryear{{Zuckerman}, {Koester}, {Reid}  \&
  {H{\"u}nsch}}{{Zuckerman} et~al.}{2003}]{Zuckerman2003}
{Zuckerman} B.,  {Koester} D.,  {Reid} I.~N.,   {H{\"u}nsch} M.,  2003, \mn@doi
  [\apj] {10.1086/377492}, \href
  {http://adsabs.harvard.edu/abs/2003ApJ...596..477Z} {596, 477}

\makeatother
\end{thebibliography}




\bsp    
\label{lastpage}
\end{document}